\documentclass[pra,aps,showpacs,floatfix,twocolumn,superscriptaddress,footinbib]{revtex4-1}
\usepackage{color}
\usepackage{colordvi}
\usepackage{graphicx,amssymb,epsfig,amsmath}
\usepackage{epstopdf}
\usepackage{amsmath}
\usepackage{bm}
\usepackage{bbold}
\usepackage{amsfonts}
\usepackage{hyperref}

\hypersetup{
    colorlinks=true,
    linkcolor=blue,
    anchorcolor=blue,
    citecolor=blue,
    filecolor=blue,      
    urlcolor=blue,
}

\usepackage{mathtools}
\usepackage{mathdots}
\usepackage{physics}

\begin{document}

\title{Interpretable and unsupervised phase classification}

\author{Julian Arnold}
\author{Frank Sch\"afer}
\affiliation{Department of Physics, University of Basel, Klingelbergstrasse 82, 4056 Basel, Switzerland}

\author{Martin \v{Z}onda}
\author{Axel U. J. Lode}
\affiliation{Institute of Physics, Albert-Ludwigs-Universit\"at Freiburg, Hermann-Herder-Strasse 3, 79104 Freiburg im Breisgau, Germany}
\date{\today}
\begin{abstract} Fully automated classification methods that yield direct physical insights into phase diagrams are of current interest. Here, we demonstrate an unsupervised machine learning method for phase classification which is rendered interpretable via an analytical derivation of its optimal predictions and allows for an automated construction scheme for order parameters. 
Based on these findings, we propose and apply an alternative, physically-motivated, data-driven scheme
which relies on the difference between mean input features. This mean-based method is computationally cheap and directly interpretable. As an example, we consider the physically rich ground-state phase diagram of the spinless Falicov-Kimball model.
\end{abstract}

\maketitle

Phase diagrams and phase transitions are of paramount importance to physics~\cite{sachdev:2011,goldenfeld:2018,carleo:2019}. 
While typical many-body systems have a large number of degrees of freedom, their phases are usually characterized by a small set of physical quantities like response functions or order parameters. However, the identification of phases and their order parameters is often a complex problem involving a large state space~\cite{sethna:2006,chaikin:1995}. 
Machine learning methods are apt for this task~\cite{carrasquilla:2017,van:2017,carrasquilla_2:2017,wang:2016,rem:2019, bohrdt:2019,dunjko:2018,ohtsuki:2017,carleo:2019,carrasquilla:2020} as they can deal with large data sets and efficiently extract information from them.
Ideally, such machine learning methods should not require any \textit{a priori} knowledge about the phases, i.e., the methods should be unsupervised~\cite{wang:2016,van:2017,van:2018,rodriguez:2019,huemeli:2018,schafer2019,greplova:2019,dawid:2020,casert:2019,wetzel:2017,che:2020,scheurer:2020,blucher:2020,zhang:2020,liu:2019,balabanov:2020,yang:2020}. Yet, they should also allow for a straightforward physical insight 
into the character of phases. Significant progress has been made recently~\cite{casert:2019, blucher:2020, zhang:2020, dawid:2020}, but some open issues with interpretability remain. Thus, unsupervised and interpretable phase classification stays a challenging, but highly rewarding task.

\begin{figure}[bth!]
\begin{center}
\includegraphics[width=0.45\textwidth]{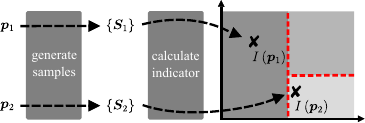}
\caption{Our workflow to predict a phase diagram with indicators $I$ for phase transitions. A set of samples $\{ \bm{S}_{i} \}$ is generated for fixed system parameters $\bm{p}_{i}$. 
Based on these samples, a scalar indicator for phase transitions, $I\left(\bm{p}_{i} \right)$, is calculated. This indicator highlights the boundaries (red) between phases (grey). 
Different unsupervised phase classification schemes are established via different indicators.}
\label{fig_fig1}
\end{center}
\end{figure}

A good example of both progress in the field and relevant issues regarding interpretability is the unsupervised method introduced in Ref.~\cite{schafer2019}. This approach is based on a predictive model trained to infer the parameters of a physical system from input data -- obtained by experimental measurements or numerical simulations -- that characterize the system's state. In the following, we refer to this approach as the \textit{prediction-based method}.
The predictions for the system parameters in the prediction-based method are changing most strongly near phase boundaries. Hence, the vector-field divergence of the deviations of the predicted system parameters from their true values serves as an indicator (label $I$ in Fig.~\ref{fig_fig1}) of phase boundaries.

The prediction-based method was hitherto successfully applied to symmetry-breaking~\cite{schafer2019}, driven-dissipative~\cite{schafer2019}, quantum~\cite{greplova:2019}, and topological phase transitions~\cite{greplova:2019,singh:2020} in various systems. The prediction-based method requires a predictive model with sufficient expressive power~\cite{bengio:2011, goodfellow:2016} to resolve different phases. The resulting phase classifications can therefore be hard to interpret if highly expressive models, such as deep neural networks (DNNs)~\cite{goodfellow:2016}, have to be used~\cite{zhang:2020, dawid:2020}. Additionally, the training of DNNs is computationally demanding. For complicated phase diagrams with a large number of phases, the applicability of the prediction-based method remains to be demonstrated.

Herein, we render the prediction-based method fully interpretable by deriving its optimal predictions.
To devise local order parameters for the predicted phases, we employ linear models~\cite{zhang:2020, blucher:2020,molnar:2019,cole:2020} to infer the system parameters. Such local order parameters distinguish neighboring phases.
As the key result of this Letter, we demonstrate a physically motivated, general, data-driven, unsupervised phase classification approach which relies on the difference between mean input features as an indicator for phase transitions (Fig.~\ref{fig_fig1}). In the following, we refer to this approach as the \textit{mean-based method}. The mean-based method eliminates the need for a predictive model, is computationally cheap, and directly interpretable.

\begin{figure*}[tbh!]
\begin{center}
\includegraphics[width=\textwidth]{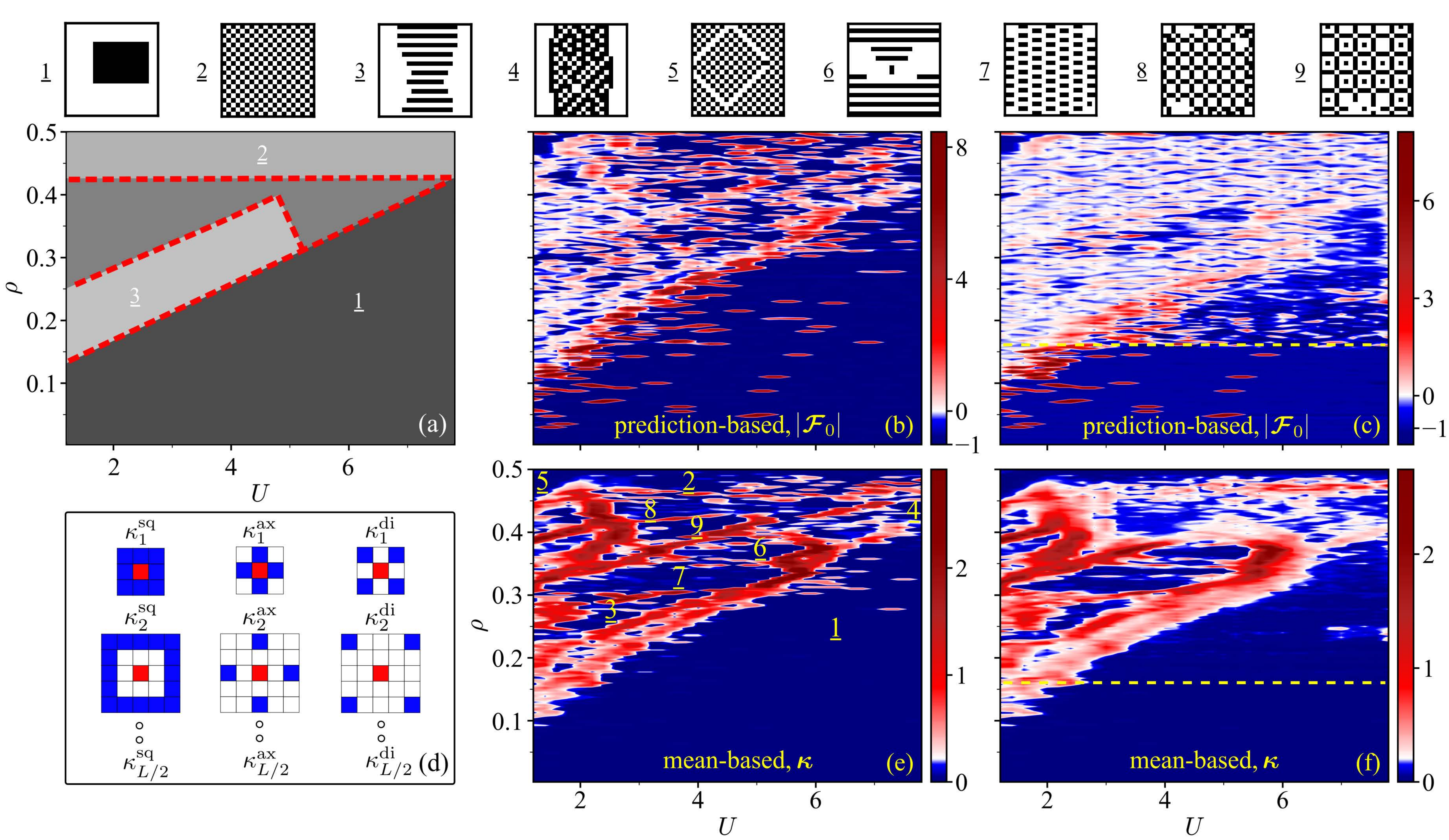}
\caption{(a) Sketch of the ground-state phase diagram of the spinless FKM. Red-dashed lines highlight the boundaries of the phases with ($\underline{1}$) segregated, ($\underline{2}$) diagonal, and ($\underline{3}$) axial orderings. For each ordering ($\underline{1}$)--($\underline{3}$), an example of a typical ground-state configuration $\bm{w}_{0}$ ($L=20$) is shown on top. Here, the absence ($w_{0,i}=0$) and presence ($w_{0,i}=1$) of an $f$ particle at lattice site $i$ is denoted by a white or black square, respectively. (b)-(c) $\nabla_{\boldsymbol{p}} \cdot \boldsymbol{\delta p}$ [Eq.~\eqref{eq_divergence}] based on the predictions of a DNN trained using $\abs{\bm{\mathcal{F}}_{0}}$ as input in the (b) noise-free and (c) noisy case. (d) Illustration of the correlation functions that measure square ($\kappa_{n}^{\rm sq}$), axial ($\kappa_{n}^{\rm ax}$), and diagonal ($\kappa_{n}^{\rm di}$) correlation at a distance $n$ from the origin [cf. Eq.~\eqref{eq_corr_main} and red and blue squares]. (e)-(f) Correlation indicator $\Delta \bar{\kappa}$ [Eq.~\eqref{eq_corrdifference}] in the (e) noise-free and (f) noisy case. Both $\nabla_{\boldsymbol{p}} \cdot \boldsymbol{\delta p}$ and $\Delta \bar{\kappa}$ serve as indicators for phase transitions (Fig.~\ref{fig_fig1}). The dashed line in (c),(f) marks the cut along $\rho=63/400\approx0.16$ analyzed in Fig.~\ref{fig_fig3}. Example configurations ($\underline{1}$)--($\underline{9}$) for some of the largest predicted regions of stability are shown on top.}
\label{fig_fig2}
\end{center}
\end{figure*}

As a physical system, we consider the two-dimensional spinless Falicov-Kimball model (FKM) \cite{FKM:1969,hubbard:1963,freericks:2003}. Its ground-state phase diagram features a large number of different phases, e.g., charge stripes or various phase separations~\cite{lemanski:2002,lemanski:2004,cencarikova:2012}. The features of these phases play an important role in the investigation of numerous physical phenomena, e.g., metal-insulator and valence transitions~\cite{plischke:1972,farkasovsky:1995a,farkasovsky:1995b,haldar:2019,kauch:2020}, pattern formations in ultracold atoms in optical lattices~\cite{maska:2008,maska:2011,hu:2015,qin:2018}, localization and correlations~\cite{maionchi:2008,antipov:2016,haldar:2017b,smith:2017,zonda:2019b,ribic:2016,ribic:2017}, 
or various nonequilibrium phenomena~\cite{freericks:2006,eckstein:2008,eckstein:2009,herrmann:2016,herrmann:2018,FreericksBook2006,zonda:2019a,smorka:2020}. Hitherto, the classification of ground-state phases in the FKM was a manual and -- due to the richness of the phase diagram~\cite{lemanski:2002,lemanski:2004,cencarikova:2012} -- lengthy and cumbersome task. The complexity of the FKM phase diagram makes it a challenging example for unsupervised and interpretable phase classification methods.

The Hamiltonian of the spinless FKM is
\begin{equation}\label{eq_fkm}
\mathcal{H}= -t\sum_{\langle ij \rangle}( d_{i}^{\dagger} d_{j}^{\phantom{\dagger}} +
 d_{j}^{\dagger} d_{i}^{\phantom{\dagger}}
)+ U \sum_{i} d_{i}^{\dagger} d_{i}^{\phantom{\dagger}} f_{i}^{\dagger} f_{i}^{\phantom{\dagger}}.
\end{equation}
Here, $t$ is the hopping integral (energy unit throughout this work), $U$ is the on-site Coulomb interaction strength, $f_{i}^{\dagger} \; (f_{i}^{\phantom{\dagger}})$ and $d_{i}^{\dagger} \; (d_{i}^{\phantom{\dagger}})$ are the creation (annihilation) operators of heavy $(f)$ and light $(d)$ fermions at lattice site $i$. The number operator $n_{f,i} = f_{i}^{\dagger} f_{i}$ commutes with the Hamiltonian for all $i$; we can replace it by its eigenvalues $w_{i} \in \{ 0,1 \}$. The ground state is thus determined by the classical $f$-particle configuration $\bm{w}=\{ w_{i} \}$ that minimizes the system energy. We focus on the ``neutral'' case~\cite{lemanski:2002}, characterized by an equal density of heavy and light particles $\rho\equiv N_f/L^2=N_d/L^2$. Here, $N_f$ ($N_d$) is the total number of heavy (light) particles and $L=20$ -- which we fix throughout this Letter -- is the linear size of the square two-dimensional lattice with periodic boundary conditions (plane symmetry group: $p4m$~\cite{doris:1978}).

Figure~\ref{fig_fig2}(a) shows a sketch of the expected phase diagram in two-dimensional parameter space~\cite{lemanski:2002}. It highlights the regions of stability of three main types of orderings, namely, ($\underline{1}$) segregated, ($\underline{2}$) diagonal, and ($\underline{3}$) axial orderings. A multitude of other phases with smaller stability regions are expected to be present in the full diagram~\cite{lemanski:2002,lemanski:2004}.

We determine the ground-state configuration $\bm{w}_{0}$ approximately for a given $\bm{p}\equiv \left(U, \rho\right)$ using an adaptive simulated annealing algorithm (Sec.~\ref{sec_SM_S1} in Supplemental Material (SM)
% SM footnote
~\footnote{See Supplemental Material for details on the simulated annealing procedure, the prediction-based, and the mean-based method, as well as a discussion of alternative phase classification methods and definitions of order parameters and correlations functions which includes Refs.~\cite{github,lemanski:2002,lemanski:2004,maska:2006,tran:2006,zonda:2012,goodfellow:2016,lecun:2012,paszke:2019,kingma:2014,blucher:2020,casert:2019, zhang:2020,scikit:2011,wetzel:2017,kreher:1999,chau:1998,wang:2016,mehta:2019,vaart:1998}.}), 
% SM footnote
where $\rho$ ranges from $1/L^2$ to half-filling $(\Delta \rho = 1/L^2)$ and $U$ ranges from $1$ to $8$ $(\Delta U=0.2)$. For each $\bm{p}$ we performed up to $64$ independent simulations. Because simulated annealing does not always converge to the ground state for large systems, we investigate two cases: a ``noise-free'' case where the best estimate $\bm{w}_0$ is taken as the ground-state and a ``noisy case'' where we take into account $10$ configurations with the lowest energies at each $\bm{p}$. The latter case is important for checking the robustness of our methods and is particularly relevant for experiments, where thermal fluctuations are inevitable.

\begin{figure*}[tbh!]
\begin{center}
\includegraphics[width=0.99\textwidth]{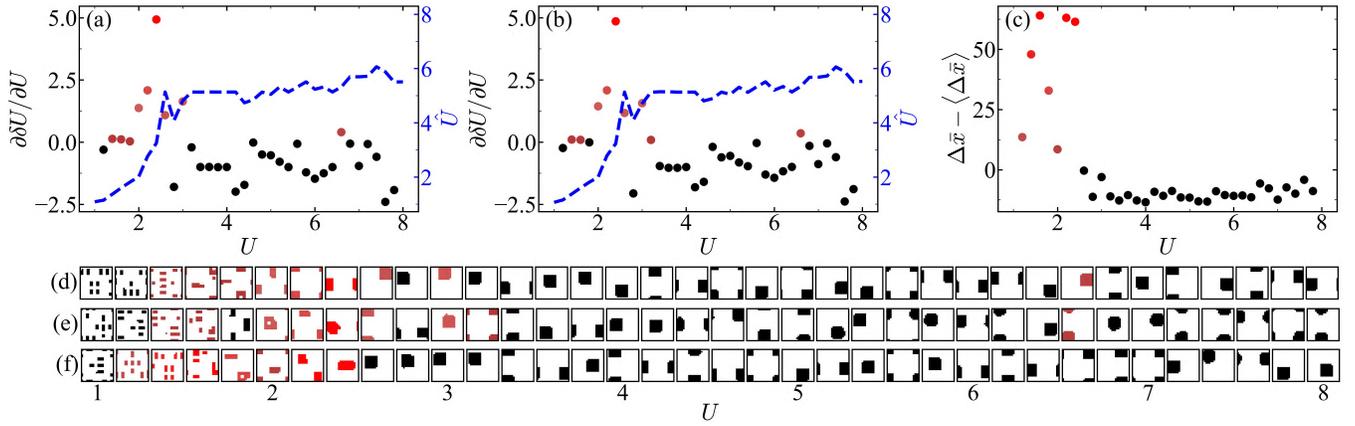}
\caption{Analysis of the transition from non-segregated to segregated orderings occurring at $U\approx2$ along the line-scan from $U_{\rm min}=1$ to $U_{\rm max}=8$ at fixed $\rho=63/400\approx0.16$. Predictions $\hat{U}$ and corresponding divergence $\partial \delta U/\partial U$ of (a),(d) a DNN and (b),(e) a linear model, as well as (c),(f) the indicator $\Delta \bar{x}$ [Eq.~\eqref{eq_differencemethod_main}] based on $\abs{\bm{\mathcal{F}}_{0}}$. $\langle \Delta \bar{x} \rangle$ denotes the average difference signal over the entire line-scan and is subtracted to account for noise arising due to finite sample statistics. The degree of red in (a)--(c) denotes an increasingly positive value of the respective indicator for phase transitions; (d)--(f) configurations visualized using the same color scale as for the points in (a)--(c), respectively.}
\label{fig_fig3}
\end{center}
\end{figure*}

We start the analysis of the phase diagram 
using the prediction-based method with DNNs (Sec.~\ref{sec_SM_S2} in SM~\cite{Note1}). 
Since the predictions $\hat{\bm{p}}\equiv\left(\hat{U}, \hat{\rho} \right)$  are most susceptible near the phase boundaries, maxima in the vector-field divergence of $\bm{\delta p} \equiv \hat{\bm{p}} - \bm{p}$ given by
\begin{equation}\label{eq_divergence}
\nabla_{\bm{p}} \cdot \bm{\delta p} = \frac{\partial \delta U}{\partial U} + \frac{\partial \delta \rho}{\partial \rho}
\end{equation}
serve as an indicator $I\left(\bm{p} \right)$ (Fig.~\ref{fig_fig1}) of phase boundaries.

We train the DNN to predict $\bm{p}=\left(U, \rho \right)$ using as input the magnitude of the two-dimensional discrete Fourier transform $\abs{\bm{\mathcal{F}}_{0}}$. Figure~\ref{fig_fig2}(b) and (c) shows the obtained phase diagram in the noise-free and the noisy case, respectively.
The usage of $\abs{\bm{\mathcal{F}}_{0}}$ instead of $\bm{w}_{0}$ results in a shorter training time because data augmentation by lattice translations is not necessary. 
Moreover, it yields an improved signal-to-noise ratio for $\nabla_{\bm{p}} \cdot \bm{\delta p}$, because the peak in $\abs{\bm{\mathcal{F}}_{0}}$ at $\bm{k}=\left(0,0\right)$ corresponds to $N_{f}$, i.e., $\rho$ is directly fed into the DNN. Consequently, the vector field $\bm{\delta p}$ exhibits a horizontal structure (Sec.~\ref{sec_SM_S6} in SM~\cite{Note1}), meaning that $\rho$ is predicted with near-perfect accuracy in both cases. The horizontal structure of $\bm{\delta p}$ implies that maxima in $\nabla_{\bm{p}} \cdot \bm{\delta p}$ indicate phase transitions along $U$ at fixed $\rho$. The $\left(U,\rho \right)$ parameter space can thus be analyzed with cuts along $U$ -- we will do this later on.

The largest connected regions with a negative divergence signal in Fig.~\ref{fig_fig2}(b) coincide with the three main regions within the sketched phase diagram displayed in Fig.~\ref{fig_fig2}(a) and their character can be confirmed 
by simple order parameter analysis (Sec.~\ref{sec_SM_S3} in SM~\cite{Note1}). Thus, the prediction-based method correctly identifies the expected main characteristics of the phase diagram. However, the phase boundaries are not reproduced with a large contrast and it is not easy to identify stability regions, besides the ones of the main orderings [$\underline{1},\underline{2},\underline{3}$ in Fig.~\ref{fig_fig2}(a)].
Specifically, the prediction-based method indicates changes of the phase at several points within stable phase regions [Fig.~\ref{fig_fig2}(b)]. These artefacts intensify in the noisy case [Fig.~\ref{fig_fig2}(c)], where -- in addition -- the training of the DNN becomes computationally heavy. Moreover, a large(er) amount of input data is needed to obtain the phase diagram with sufficient accuracy.
To cope with these problems, it is first necessary to understand the DNN predictions.

For this purpose, we derive the form of the optimal predictive model for the prediction-based method.
In the general noisy case, the optimal prediction for an input $\bm{x}$ of a model trained to minimize a mean-square-error loss (derivation in Sec.~\ref{sec_SM_S2} in SM~\cite{Note1}) is 
\begin{equation}\label{eq_noisy_optimalpredictions_main}
\hat{\bm{p}}_{\rm opt}\left( \bm{x} \right) = \frac{\sum_{i} {\rm P}_{i}\left( \bm{x}\right) \bm{p}_{i}}{ \sum_{i} {\rm P}_{i}\left(\bm{x}\right)}.
\end{equation}
Here, the sum runs over all sampled points $\{\bm{p}_{i} \}$ in parameter space. The probability of drawing the input $\bm{x}$ at $\bm{p}_{i}$ is governed by the distribution ${\rm P}_{i}$. By considering identical, non-zero values of ${\rm P}_{i}\left(\bm{x}\right)$ within a particular region of parameter space and zero outside, we get the optimal predictions for the noise-free case. The vector-field divergence $\nabla_{\bm{p}} \cdot \bm{\delta p}$ 
[Fig.~\ref{fig_fig2}(b)] matches the one of an optimal predictive model. We, therefore, successfully rendered the phase classification of the prediction-based method interpretable. However, we still lack physical insights into the character of the predicted phases.

An important observation concerning the noise-free case is that the method predicts a phase transition whenever neighboring configurations in $U$ (at $\rho=\text{const.}$) are not related by transformations of $p4m$. The optimal model predictions for all points within such a phase are placed at its center of mass. This results in a signal of $\nabla_{\boldsymbol{p}} \cdot \boldsymbol{\delta p} = -1$ within a phase.  
The value of $\nabla_{\bm{p}} \cdot \bm{\delta p}$ at two points $\bm{p}$ that constitute a phase boundary serves as a measure of the mean extent of the two corresponding phases (along $U$). Consequently, the large extent of the segregated phase in parameter space is the cause for the large, isolated maxima in $\nabla_{\boldsymbol{p}} \cdot \boldsymbol{\delta p}$ within it, Fig.~\ref{fig_fig2}(b). Such isolated points are physically meaningless.

The noisy case can be understood from a single line-scan. At $\rho=63/400$ [dashed line in Fig.~\ref{fig_fig2}(c)] 
a transition from a non-segregated to a segregated ordering occurs. Figure~\ref{fig_fig3}(a),(d) shows that the predictions $\hat{U}$ and the corresponding divergence $\partial \delta U/\partial U$ obtained with a DNN indicate the corresponding phase boundary. 
Finite-sample statistics cause significant fluctuations in the distributions ${\rm P}_{i}\left(\abs{\bm{\mathcal{F}}_{0}} \right)$ 
and varying predictions $\hat{U}$ within the segregated phase. 
The fluctuations in ${\rm P}_{i}$ can yield divergence signals close to zero that, again, correspond to misleading predictions of phase boundaries within the segregated phase.

Nevertheless, we can now can use 
the predictions $\hat{U}$ for the definition of automatically generated local order parameters. 
The example in Fig.~\ref{fig_fig3}(a) exhibits 
the largest change in $\hat{U}$ at the transition from non-segregated to segregated orderings; $\hat{U}$ increases only slowly within the segregated phase ($U\gtrsim3$). Moreover, the predictions $\hat{U}$ can be qualitatively reproduced by training a linear model as opposed to a DNN [Fig.~\ref{fig_fig3}(b),(e)]. The local order parameter $\hat{U}$ based on the linear model is directly interpretable through its weights and biases. This approach also works in the ``noise-free'' case and with other inputs
(Sec.~\ref{sec_SM_S6} in SM~\cite{Note1}). 
We thus demonstrate that the construction of local order parameters can be automated by training a linear model for each individual phase transition which was identified prior using a DNN with the prediction-based method. Eventually, the DNN-based model can be replaced by the corresponding set of linear models which is conceptually related to training local surrogate models~\cite{molnar:2019}.

Importantly, the form of the optimal model in the prediction-based method paves the way for a class of computationally cheap algorithms for unsupervised phase classification without any predictive model. Here, we focus on the \textit{mean-based method} (other methods in Sec.~\ref{sec_SM_S4} in SM~\cite{Note1}).
The optimal model predictions reveal that the corresponding predicted phase diagram can be reproduced by detecting changes in observables derived from $\bm{w}_{0}$ which are invariant under transformations of $p4m$.
Ideally, the observable should not be very sensitive to small changes in $\bm{w}_{0}$ within a stability region.

A suitable physically-motivated choice are correlation functions that measure the order of a given configuration: 
\begin{align}\label{eq_corr_main}
\kappa_{n}^{\xi}\left(U,\rho \right) &= \frac{1}{m^\xi_nL^2} \sum_{i=1}^{L^2} \sum_{j\in{\{j^\xi_n\}}}\left(2w_i-1\right)\left(2w_j-1\right),
\end{align}
where $m^\xi_n$ is the number of constituents in the set $\{j^\xi_n\}$
which contains lattice points matching three types of orders that measure square ($\xi=\rm{sq}$), axial ($\xi=\rm{ax}$), and diagonal ($\xi=\rm{di}$) correlations over $n$ lattice sites (illustration in Fig.~\ref{fig_fig2}(d) and details in Sec.~\ref{sec_SM_S3} in SM~\cite{Note1}).

With these correlation functions, we define the following {\it correlation indicator}
for the mean-based method:
\begin{equation}\label{eq_corrdifference}
\Delta \bar{\kappa} \left( U, \rho \right) \equiv \norm{\bar{\bm{\kappa}}\left(U+\Delta U, \rho\right) - \bar{\bm{\kappa}}\left(U-\Delta U, \rho\right)},
\end{equation}
at each point $\bm{p}=\left(U,\rho \right)$, where $\bar{\bm{\kappa}}=[\bar{\kappa}_{1}^{\rm sq},...,\bar{\kappa}_{L/2}^{\rm sq}, \bar{\kappa}_{1}^{\rm ax},...,\bar{\kappa}_{L/2}^{\rm ax},\bar{\kappa}_{1}^{\rm di},...,\bar{\kappa}_{L/2}^{\rm di}]^{\rm T}$. Here, the $\bar{\cdot}$-notation indicates the average over all inputs at a given point $\bm{p}$, if multiple inputs are considered.
The indicator $\Delta \bar{\kappa}$ measures the magnitude of the change of order quantified by $\kappa_{n}^{\xi}$.

Our results for, both, noisy and noise-free cases 
demonstrate that the mean-based method with the indicator $\Delta \bar{\kappa}$ 
reveals the phase diagram more clearly than $\nabla_{\bm{p}} \cdot \bm{\delta p}$, compare Fig.~\ref{fig_fig2}(e),(f) and (b,c). The 
indicator $\Delta \bar{\kappa}$
reproduces the main characteristics of the FKM phase diagram [Fig.~\ref{fig_fig2}(a)]: $\Delta \bar{\kappa}$ almost vanishes within the stability region of segregated orderings, in the presence or absence of noise [Fig.~\ref{fig_fig2}(e),(f)], and marks all phase boundaries of Fig.~\ref{fig_fig2}(a). Moreover, we obtain a detailed, physical subdivision of the phase diagram -- see the identified orderings in Fig.~\ref{fig_fig2}(top) and labels ($\underline{1}$)--($\underline{9}$) in Fig.~\ref{fig_fig2}(e).
We infer that the mean-based method with the correlation indicator $\Delta \bar{\kappa}$ in Eq.~\eqref{eq_corrdifference} is an excellent tool to detect phase boundaries.

We now ask the question if the mean-based method can be applied to the phase classification problem without a specific physically-motivated input. To this end, we extend the mean-based method to general inputs, by defining a {\it generic indicator}:
\begin{equation}\label{eq_differencemethod_main}
\Delta \bar{x} \left( \bm{p} \right) \equiv \norm{\bar{\bm{x}}\left(U+\Delta U, \rho\right) - \bar{\bm{x}}\left(U-\Delta U, \rho\right)}.
\end{equation}
Here, $\bar{\bm{x}}\left(\bm{p}_{i} \right)= \sum_{j} {\rm P}_{i}\left( \bm{x}_{j} \right) \bm{x}_{j} \approx 1/N \sum_{j} \bm{x}_{j}\left(\bm{p}_{i} \right)$ denotes the average input at a point $\bm{p}_{i}$ over all corresponding $N$ inputs $\bm{x}_{j}$ (Sec.~\ref{sec_SM_S5} in SM~\cite{Note1}).

Since the generic indicator $\Delta \bar{x}$
detects phase transitions, Eq.~\eqref{eq_differencemethod_main} establishes a general, data-driven scheme without a predictive model which is applicable in both the noise-free and noisy case.
In any such data-driven approach the used input $\bm{x}$ 
crucially affects the performance of the phase classification; Fig.~\ref{fig_fig2}(e),(f) show that the set of correlation functions $\bm{\kappa}$ are an appropriate choice in the case of the FKM (results with different inputs in Sec.~\ref{sec_SM_S6} in SM~\cite{Note1}).

Another suitable choice for the input for the mean-based method are the Fourier transformed configurations $\abs{\bm{\mathcal{F}}_{0}}$ [line-scan along $U$ in Fig.~\ref{fig_fig3}(c),(f)]. Even in this case the difference signal is a good indicator for the phase transition at $U\approx2$. This underpins the generality and robustness of the mean-based method and shows its possible applicability to other models beyond the FKM, where the inputs are generally different.    

We stress that the indicators of phase transitions in the mean-based method [Eq.~\eqref{eq_differencemethod_main}] and the prediction-based method [Eq.~\eqref{eq_divergence}] differ fundamentally. They constitute two distinct approaches to characterize changes in the underlying distributions ${\rm P}\left(\bm{x}\right)$. However, the mean-based method has some clear advantages: it is computationally cheap and allows for direct physical insights. For example, by calculating an indicator, Eq.~\eqref{eq_differencemethod_main}, based on each individual element of $\bm{\kappa}$ or $ \abs{\bm{\mathcal{F}}_{0}}$ one can directly find the elements which have the greatest impact on the overall indicator. This further characterizes each phase transition and simplifies the subsequent analysis.

In conclusion, we have rendered the prediction-based method fully interpretable with a derivation of its optimal model predictions and the automatic local-order-parameter generation using linear surrogate models. Moreover, we have presented a mean-based method that works outstandingly well as an unsupervised phase classification approach for various inputs and in the presence of noise. We infer that applications of our mean-based method to arbitrary phase diagrams featuring, e.g., quantum or topological phase transitions are feasible. Specifically, applications to quantum-classical systems such as the FKM and its numerous generalizations~\cite{freericks:2006,cencarikova:2012,petrovic:2018,li:2019,goncalves:2019} are now straightforward. The success of the mean-based method suggests extensions to unsupervised phase classification methods based on higher-order moments or alternative distance measures.

We would like to thank Niels L\"orch, Eliska Greplova, Michael Thoss, and Christoph Bruder for inspiring discussions. J.A. and F.S. acknowledge financial support from the Swiss National Science Foundation (SNSF) and the NCCR Quantum Science and Technology. A.U.J.L. acknowledges financial support by the Austrian Science Foundation (FWF) under grant No. P-32033-N32. Computation time on the Hawk cluster at the HLRS Stuttgart and at sciCORE (scicore.unibas.ch) scientific computing core facility at University of Basel, as well as support by the state of Baden-W\"urttemberg through bwHPC and the German Research Foundation (DFG) through grants no INST 40/467-1 FUGG (JUSTUS cluster), INST 39/963-1 FUGG (bwForCluster NEMO), and INST 37/935-1 FUGG (bwForCluster BinAC) is gratefully acknowledged.

\let\oldaddcontentsline\addcontentsline% Store \addcontentsline
\renewcommand{\addcontentsline}[3]{}% Make \addcontentsline a no-op
\bibliography{refs}

\begin{thebibliography}{81}%
\makeatletter
\providecommand \@ifxundefined [1]{%
 \@ifx{#1\undefined}
}%
\providecommand \@ifnum [1]{%
 \ifnum #1\expandafter \@firstoftwo
 \else \expandafter \@secondoftwo
 \fi
}%
\providecommand \@ifx [1]{%
 \ifx #1\expandafter \@firstoftwo
 \else \expandafter \@secondoftwo
 \fi
}%
\providecommand \natexlab [1]{#1}%
\providecommand \enquote  [1]{``#1''}%
\providecommand \bibnamefont  [1]{#1}%
\providecommand \bibfnamefont [1]{#1}%
\providecommand \citenamefont [1]{#1}%
\providecommand \href@noop [0]{\@secondoftwo}%
\providecommand \href [0]{\begingroup \@sanitize@url \@href}%
\providecommand \@href[1]{\@@startlink{#1}\@@href}%
\providecommand \@@href[1]{\endgroup#1\@@endlink}%
\providecommand \@sanitize@url [0]{\catcode `\\12\catcode `\$12\catcode
  `\&12\catcode `\#12\catcode `\^12\catcode `\_12\catcode `\%12\relax}%
\providecommand \@@startlink[1]{}%
\providecommand \@@endlink[0]{}%
\providecommand \url  [0]{\begingroup\@sanitize@url \@url }%
\providecommand \@url [1]{\endgroup\@href {#1}{\urlprefix }}%
\providecommand \urlprefix  [0]{URL }%
\providecommand \Eprint [0]{\href }%
\providecommand \doibase [0]{http://dx.doi.org/}%
\providecommand \selectlanguage [0]{\@gobble}%
\providecommand \bibinfo  [0]{\@secondoftwo}%
\providecommand \bibfield  [0]{\@secondoftwo}%
\providecommand \translation [1]{[#1]}%
\providecommand \BibitemOpen [0]{}%
\providecommand \bibitemStop [0]{}%
\providecommand \bibitemNoStop [0]{.\EOS\space}%
\providecommand \EOS [0]{\spacefactor3000\relax}%
\providecommand \BibitemShut  [1]{\csname bibitem#1\endcsname}%
\let\auto@bib@innerbib\@empty
%</preamble>
\bibitem [{\citenamefont {Sachdev}(2011)}]{sachdev:2011}%
  \BibitemOpen
  \bibfield  {author} {\bibinfo {author} {\bibfnamefont {S.}~\bibnamefont
  {Sachdev}},\ }\href {\doibase 10.1017/cbo9780511973765} {\emph {\bibinfo
  {title} {Quantum Phase Transitions}}}\ (\bibinfo  {publisher} {Cambridge
  University Press},\ \bibinfo {year} {2011})\BibitemShut {NoStop}%
\bibitem [{\citenamefont {Goldenfeld}(2018)}]{goldenfeld:2018}%
  \BibitemOpen
  \bibfield  {author} {\bibinfo {author} {\bibfnamefont {N.}~\bibnamefont
  {Goldenfeld}},\ }\href {\doibase 10.1201/9780429493492} {\emph {\bibinfo
  {title} {Lectures On Phase Transitions And The Renormalization Group}}}\
  (\bibinfo  {publisher} {CRC Press},\ \bibinfo {year} {2018})\BibitemShut
  {NoStop}%
\bibitem [{\citenamefont {Carleo}\ \emph {et~al.}(2019)\citenamefont {Carleo},
  \citenamefont {Cirac}, \citenamefont {Cranmer}, \citenamefont {Daudet},
  \citenamefont {Schuld}, \citenamefont {Tishby}, \citenamefont
  {Vogt-Maranto},\ and\ \citenamefont {Zdeborov\'a}}]{carleo:2019}%
  \BibitemOpen
  \bibfield  {author} {\bibinfo {author} {\bibfnamefont {G.}~\bibnamefont
  {Carleo}}, \bibinfo {author} {\bibfnamefont {I.}~\bibnamefont {Cirac}},
  \bibinfo {author} {\bibfnamefont {K.}~\bibnamefont {Cranmer}}, \bibinfo
  {author} {\bibfnamefont {L.}~\bibnamefont {Daudet}}, \bibinfo {author}
  {\bibfnamefont {M.}~\bibnamefont {Schuld}}, \bibinfo {author} {\bibfnamefont
  {N.}~\bibnamefont {Tishby}}, \bibinfo {author} {\bibfnamefont
  {L.}~\bibnamefont {Vogt-Maranto}}, \ and\ \bibinfo {author} {\bibfnamefont
  {L.}~\bibnamefont {Zdeborov\'a}},\ }\href {\doibase
  10.1103/RevModPhys.91.045002} {\bibfield  {journal} {\bibinfo  {journal}
  {Rev. Mod. Phys.}\ }\textbf {\bibinfo {volume} {91}},\ \bibinfo {pages}
  {045002} (\bibinfo {year} {2019})}\BibitemShut {NoStop}%
\bibitem [{\citenamefont {Sethna}(2006)}]{sethna:2006}%
  \BibitemOpen
  \bibfield  {author} {\bibinfo {author} {\bibfnamefont {J.}~\bibnamefont
  {Sethna}},\ }\href
  {https://global.oup.com/academic/product/statistical-mechanics-9780198566779?cc=ch&lang=en&}
  {\emph {\bibinfo {title} {Statistical {M}echanics: {E}ntropy, {O}rder
  {P}arameters, and {C}omplexity}}}\ (\bibinfo  {publisher} {Oxford University
  Press},\ \bibinfo {year} {2006})\BibitemShut {NoStop}%
\bibitem [{\citenamefont {Chaikin}\ and\ \citenamefont
  {Lubensky}(1995)}]{chaikin:1995}%
  \BibitemOpen
  \bibfield  {author} {\bibinfo {author} {\bibfnamefont {P.~M.}\ \bibnamefont
  {Chaikin}}\ and\ \bibinfo {author} {\bibfnamefont {T.~C.}\ \bibnamefont
  {Lubensky}},\ }\href {\doibase 10.1017/CBO9780511813467} {\emph {\bibinfo
  {title} {Principles of Condensed Matter Physics}}}\ (\bibinfo  {publisher}
  {Cambridge University Press},\ \bibinfo {year} {1995})\BibitemShut {NoStop}%
\bibitem [{\citenamefont {Carrasquilla}\ and\ \citenamefont
  {Melko}(2017)}]{carrasquilla:2017}%
  \BibitemOpen
  \bibfield  {author} {\bibinfo {author} {\bibfnamefont {J.}~\bibnamefont
  {Carrasquilla}}\ and\ \bibinfo {author} {\bibfnamefont {R.~G.}\ \bibnamefont
  {Melko}},\ }\href {\doibase 10.1038/nphys4035} {\bibfield  {journal}
  {\bibinfo  {journal} {Nat. Phys.}\ }\textbf {\bibinfo {volume} {13}},\
  \bibinfo {pages} {431} (\bibinfo {year} {2017})}\BibitemShut {NoStop}%
\bibitem [{\citenamefont {Van~Nieuwenburg}\ \emph {et~al.}(2017)\citenamefont
  {Van~Nieuwenburg}, \citenamefont {Liu},\ and\ \citenamefont
  {Huber}}]{van:2017}%
  \BibitemOpen
  \bibfield  {author} {\bibinfo {author} {\bibfnamefont {E.~P.}\ \bibnamefont
  {Van~Nieuwenburg}}, \bibinfo {author} {\bibfnamefont {Y.-H.}\ \bibnamefont
  {Liu}}, \ and\ \bibinfo {author} {\bibfnamefont {S.~D.}\ \bibnamefont
  {Huber}},\ }\href {\doibase 10.1038/nphys4037} {\bibfield  {journal}
  {\bibinfo  {journal} {Nat. Phys.}\ }\textbf {\bibinfo {volume} {13}},\
  \bibinfo {pages} {435} (\bibinfo {year} {2017})}\BibitemShut {NoStop}%
\bibitem [{\citenamefont {Ch'ng}\ \emph {et~al.}(2017)\citenamefont {Ch'ng},
  \citenamefont {Carrasquilla}, \citenamefont {Melko},\ and\ \citenamefont
  {Khatami}}]{carrasquilla_2:2017}%
  \BibitemOpen
  \bibfield  {author} {\bibinfo {author} {\bibfnamefont {K.}~\bibnamefont
  {Ch'ng}}, \bibinfo {author} {\bibfnamefont {J.}~\bibnamefont {Carrasquilla}},
  \bibinfo {author} {\bibfnamefont {R.~G.}\ \bibnamefont {Melko}}, \ and\
  \bibinfo {author} {\bibfnamefont {E.}~\bibnamefont {Khatami}},\ }\href
  {\doibase 10.1103/PhysRevX.7.031038} {\bibfield  {journal} {\bibinfo
  {journal} {Phys. Rev. X}\ }\textbf {\bibinfo {volume} {7}},\ \bibinfo {pages}
  {031038} (\bibinfo {year} {2017})}\BibitemShut {NoStop}%
\bibitem [{\citenamefont {Wang}(2016)}]{wang:2016}%
  \BibitemOpen
  \bibfield  {author} {\bibinfo {author} {\bibfnamefont {L.}~\bibnamefont
  {Wang}},\ }\href {\doibase 10.1103/PhysRevB.94.195105} {\bibfield  {journal}
  {\bibinfo  {journal} {Phys. Rev. B}\ }\textbf {\bibinfo {volume} {94}},\
  \bibinfo {pages} {195105} (\bibinfo {year} {2016})}\BibitemShut {NoStop}%
\bibitem [{\citenamefont {Rem}\ \emph {et~al.}(2019)\citenamefont {Rem},
  \citenamefont {K{\"a}ming}, \citenamefont {Tarnowski}, \citenamefont
  {Asteria}, \citenamefont {Fl{\"a}schner}, \citenamefont {Becker},
  \citenamefont {Sengstock},\ and\ \citenamefont {Weitenberg}}]{rem:2019}%
  \BibitemOpen
  \bibfield  {author} {\bibinfo {author} {\bibfnamefont {B.~S.}\ \bibnamefont
  {Rem}}, \bibinfo {author} {\bibfnamefont {N.}~\bibnamefont {K{\"a}ming}},
  \bibinfo {author} {\bibfnamefont {M.}~\bibnamefont {Tarnowski}}, \bibinfo
  {author} {\bibfnamefont {L.}~\bibnamefont {Asteria}}, \bibinfo {author}
  {\bibfnamefont {N.}~\bibnamefont {Fl{\"a}schner}}, \bibinfo {author}
  {\bibfnamefont {C.}~\bibnamefont {Becker}}, \bibinfo {author} {\bibfnamefont
  {K.}~\bibnamefont {Sengstock}}, \ and\ \bibinfo {author} {\bibfnamefont
  {C.}~\bibnamefont {Weitenberg}},\ }\href {\doibase 10.1038/s41567-019-0554-0}
  {\bibfield  {journal} {\bibinfo  {journal} {Nat. Phys.}\ }\textbf {\bibinfo
  {volume} {15}},\ \bibinfo {pages} {917} (\bibinfo {year} {2019})}\BibitemShut
  {NoStop}%
\bibitem [{\citenamefont {Bohrdt}\ \emph {et~al.}(2019)\citenamefont {Bohrdt},
  \citenamefont {Chiu}, \citenamefont {Ji}, \citenamefont {Xu}, \citenamefont
  {Greif}, \citenamefont {Greiner}, \citenamefont {Demler}, \citenamefont
  {Grusdt},\ and\ \citenamefont {Knap}}]{bohrdt:2019}%
  \BibitemOpen
  \bibfield  {author} {\bibinfo {author} {\bibfnamefont {A.}~\bibnamefont
  {Bohrdt}}, \bibinfo {author} {\bibfnamefont {C.~S.}\ \bibnamefont {Chiu}},
  \bibinfo {author} {\bibfnamefont {G.}~\bibnamefont {Ji}}, \bibinfo {author}
  {\bibfnamefont {M.}~\bibnamefont {Xu}}, \bibinfo {author} {\bibfnamefont
  {D.}~\bibnamefont {Greif}}, \bibinfo {author} {\bibfnamefont
  {M.}~\bibnamefont {Greiner}}, \bibinfo {author} {\bibfnamefont
  {E.}~\bibnamefont {Demler}}, \bibinfo {author} {\bibfnamefont
  {F.}~\bibnamefont {Grusdt}}, \ and\ \bibinfo {author} {\bibfnamefont
  {M.}~\bibnamefont {Knap}},\ }\href {\doibase 10.1038/s41567-019-0565-x}
  {\bibfield  {journal} {\bibinfo  {journal} {Nat. Phys.}\ }\textbf {\bibinfo
  {volume} {15}},\ \bibinfo {pages} {921} (\bibinfo {year} {2019})}\BibitemShut
  {NoStop}%
\bibitem [{\citenamefont {Dunjko}\ and\ \citenamefont
  {Briegel}(2018)}]{dunjko:2018}%
  \BibitemOpen
  \bibfield  {author} {\bibinfo {author} {\bibfnamefont {V.}~\bibnamefont
  {Dunjko}}\ and\ \bibinfo {author} {\bibfnamefont {H.~J.}\ \bibnamefont
  {Briegel}},\ }\href {\doibase 10.1088/1361-6633/aab406} {\bibfield  {journal}
  {\bibinfo  {journal} {Reports on Progress in Physics}\ }\textbf {\bibinfo
  {volume} {81}},\ \bibinfo {pages} {074001} (\bibinfo {year}
  {2018})}\BibitemShut {NoStop}%
\bibitem [{\citenamefont {Ohtsuki}\ and\ \citenamefont
  {Ohtsuki}(2017)}]{ohtsuki:2017}%
  \BibitemOpen
  \bibfield  {author} {\bibinfo {author} {\bibfnamefont {T.}~\bibnamefont
  {Ohtsuki}}\ and\ \bibinfo {author} {\bibfnamefont {T.}~\bibnamefont
  {Ohtsuki}},\ }\href {\doibase 10.7566/JPSJ.86.044708} {\bibfield  {journal}
  {\bibinfo  {journal} {Journal of the Physical Society of Japan}\ }\textbf
  {\bibinfo {volume} {86}},\ \bibinfo {pages} {044708} (\bibinfo {year}
  {2017})}\BibitemShut {NoStop}%
\bibitem [{\citenamefont {Carrasquilla}(2020)}]{carrasquilla:2020}%
  \BibitemOpen
  \bibfield  {author} {\bibinfo {author} {\bibfnamefont {J.}~\bibnamefont
  {Carrasquilla}},\ }\href {\doibase 10.1080/23746149.2020.1797528} {\bibfield
  {journal} {\bibinfo  {journal} {Advances in Physics: X}\ }\textbf {\bibinfo
  {volume} {5}},\ \bibinfo {pages} {1797528} (\bibinfo {year}
  {2020})}\BibitemShut {NoStop}%
\bibitem [{\citenamefont {Liu}\ and\ \citenamefont {van
  Nieuwenburg}(2018)}]{van:2018}%
  \BibitemOpen
  \bibfield  {author} {\bibinfo {author} {\bibfnamefont {Y.-H.}\ \bibnamefont
  {Liu}}\ and\ \bibinfo {author} {\bibfnamefont {E.~P.~L.}\ \bibnamefont {van
  Nieuwenburg}},\ }\href {\doibase 10.1103/PhysRevLett.120.176401} {\bibfield
  {journal} {\bibinfo  {journal} {Phys. Rev. Lett.}\ }\textbf {\bibinfo
  {volume} {120}},\ \bibinfo {pages} {176401} (\bibinfo {year}
  {2018})}\BibitemShut {NoStop}%
\bibitem [{\citenamefont {Rodriguez-Nieva}\ and\ \citenamefont
  {Scheurer}(2019)}]{rodriguez:2019}%
  \BibitemOpen
  \bibfield  {author} {\bibinfo {author} {\bibfnamefont {J.~F.}\ \bibnamefont
  {Rodriguez-Nieva}}\ and\ \bibinfo {author} {\bibfnamefont {M.~S.}\
  \bibnamefont {Scheurer}},\ }\href {\doibase 10.1038/s41567-019-0512-x}
  {\bibfield  {journal} {\bibinfo  {journal} {Nat. Phys.}\ }\textbf {\bibinfo
  {volume} {15}},\ \bibinfo {pages} {790} (\bibinfo {year} {2019})}\BibitemShut
  {NoStop}%
\bibitem [{\citenamefont {Huembeli}\ \emph {et~al.}(2018)\citenamefont
  {Huembeli}, \citenamefont {Dauphin},\ and\ \citenamefont
  {Wittek}}]{huemeli:2018}%
  \BibitemOpen
  \bibfield  {author} {\bibinfo {author} {\bibfnamefont {P.}~\bibnamefont
  {Huembeli}}, \bibinfo {author} {\bibfnamefont {A.}~\bibnamefont {Dauphin}}, \
  and\ \bibinfo {author} {\bibfnamefont {P.}~\bibnamefont {Wittek}},\ }\href
  {\doibase 10.1103/PhysRevB.97.134109} {\bibfield  {journal} {\bibinfo
  {journal} {Phys. Rev. B}\ }\textbf {\bibinfo {volume} {97}},\ \bibinfo
  {pages} {134109} (\bibinfo {year} {2018})}\BibitemShut {NoStop}%
\bibitem [{\citenamefont {Sch\"afer}\ and\ \citenamefont
  {L\"orch}(2019)}]{schafer2019}%
  \BibitemOpen
  \bibfield  {author} {\bibinfo {author} {\bibfnamefont {F.}~\bibnamefont
  {Sch\"afer}}\ and\ \bibinfo {author} {\bibfnamefont {N.}~\bibnamefont
  {L\"orch}},\ }\href {\doibase 10.1103/PhysRevE.99.062107} {\bibfield
  {journal} {\bibinfo  {journal} {Phys. Rev. E}\ }\textbf {\bibinfo {volume}
  {99}},\ \bibinfo {pages} {062107} (\bibinfo {year} {2019})}\BibitemShut
  {NoStop}%
\bibitem [{\citenamefont {Greplova}\ \emph {et~al.}(2020)\citenamefont
  {Greplova}, \citenamefont {Valenti}, \citenamefont {Boschung}, \citenamefont
  {Schäfer}, \citenamefont {Lörch},\ and\ \citenamefont
  {Huber}}]{greplova:2019}%
  \BibitemOpen
  \bibfield  {author} {\bibinfo {author} {\bibfnamefont {E.}~\bibnamefont
  {Greplova}}, \bibinfo {author} {\bibfnamefont {A.}~\bibnamefont {Valenti}},
  \bibinfo {author} {\bibfnamefont {G.}~\bibnamefont {Boschung}}, \bibinfo
  {author} {\bibfnamefont {F.}~\bibnamefont {Schäfer}}, \bibinfo {author}
  {\bibfnamefont {N.}~\bibnamefont {Lörch}}, \ and\ \bibinfo {author}
  {\bibfnamefont {S.~D.}\ \bibnamefont {Huber}},\ }\href {\doibase
  10.1088/1367-2630/ab7771} {\bibfield  {journal} {\bibinfo  {journal} {New
  Journal of Physics}\ }\textbf {\bibinfo {volume} {22}},\ \bibinfo {pages}
  {045003} (\bibinfo {year} {2020})}\BibitemShut {NoStop}%
\bibitem [{\citenamefont {Dawid}\ \emph {et~al.}(2020)\citenamefont {Dawid},
  \citenamefont {Huembeli}, \citenamefont {Tomza}, \citenamefont {Lewenstein},\
  and\ \citenamefont {Dauphin}}]{dawid:2020}%
  \BibitemOpen
  \bibfield  {author} {\bibinfo {author} {\bibfnamefont {A.}~\bibnamefont
  {Dawid}}, \bibinfo {author} {\bibfnamefont {P.}~\bibnamefont {Huembeli}},
  \bibinfo {author} {\bibfnamefont {M.}~\bibnamefont {Tomza}}, \bibinfo
  {author} {\bibfnamefont {M.}~\bibnamefont {Lewenstein}}, \ and\ \bibinfo
  {author} {\bibfnamefont {A.}~\bibnamefont {Dauphin}},\ }\href
  {https://arxiv.org/abs/2004.04711} {\bibfield  {journal} {\bibinfo  {journal}
  {arXiv:2004.04711}\ } (\bibinfo {year} {2020})}\BibitemShut {NoStop}%
\bibitem [{\citenamefont {Casert}\ \emph {et~al.}(2019)\citenamefont {Casert},
  \citenamefont {Vieijra}, \citenamefont {Nys},\ and\ \citenamefont
  {Ryckebusch}}]{casert:2019}%
  \BibitemOpen
  \bibfield  {author} {\bibinfo {author} {\bibfnamefont {C.}~\bibnamefont
  {Casert}}, \bibinfo {author} {\bibfnamefont {T.}~\bibnamefont {Vieijra}},
  \bibinfo {author} {\bibfnamefont {J.}~\bibnamefont {Nys}}, \ and\ \bibinfo
  {author} {\bibfnamefont {J.}~\bibnamefont {Ryckebusch}},\ }\href {\doibase
  10.1103/PhysRevE.99.023304} {\bibfield  {journal} {\bibinfo  {journal} {Phys.
  Rev. E}\ }\textbf {\bibinfo {volume} {99}},\ \bibinfo {pages} {023304}
  (\bibinfo {year} {2019})}\BibitemShut {NoStop}%
\bibitem [{\citenamefont {Wetzel}(2017)}]{wetzel:2017}%
  \BibitemOpen
  \bibfield  {author} {\bibinfo {author} {\bibfnamefont {S.~J.}\ \bibnamefont
  {Wetzel}},\ }\href {\doibase 10.1103/PhysRevE.96.022140} {\bibfield
  {journal} {\bibinfo  {journal} {Phys. Rev. E}\ }\textbf {\bibinfo {volume}
  {96}},\ \bibinfo {pages} {022140} (\bibinfo {year} {2017})}\BibitemShut
  {NoStop}%
\bibitem [{\citenamefont {Che}\ \emph {et~al.}(2020)\citenamefont {Che},
  \citenamefont {Gneiting}, \citenamefont {Liu},\ and\ \citenamefont
  {Nori}}]{che:2020}%
  \BibitemOpen
  \bibfield  {author} {\bibinfo {author} {\bibfnamefont {Y.}~\bibnamefont
  {Che}}, \bibinfo {author} {\bibfnamefont {C.}~\bibnamefont {Gneiting}},
  \bibinfo {author} {\bibfnamefont {T.}~\bibnamefont {Liu}}, \ and\ \bibinfo
  {author} {\bibfnamefont {F.}~\bibnamefont {Nori}},\ }\href
  {https://arxiv.org/abs/2002.02363} {\bibfield  {journal} {\bibinfo  {journal}
  {arXiv:2002.02363}\ } (\bibinfo {year} {2020})}\BibitemShut {NoStop}%
\bibitem [{\citenamefont {Scheurer}\ and\ \citenamefont
  {Slager}(2020)}]{scheurer:2020}%
  \BibitemOpen
  \bibfield  {author} {\bibinfo {author} {\bibfnamefont {M.~S.}\ \bibnamefont
  {Scheurer}}\ and\ \bibinfo {author} {\bibfnamefont {R.-J.}\ \bibnamefont
  {Slager}},\ }\href {\doibase 10.1103/PhysRevLett.124.226401} {\bibfield
  {journal} {\bibinfo  {journal} {Phys. Rev. Lett.}\ }\textbf {\bibinfo
  {volume} {124}},\ \bibinfo {pages} {226401} (\bibinfo {year}
  {2020})}\BibitemShut {NoStop}%
\bibitem [{\citenamefont {Bl\"ucher}\ \emph {et~al.}(2020)\citenamefont
  {Bl\"ucher}, \citenamefont {Kades}, \citenamefont {Pawlowski}, \citenamefont
  {Strodthoff},\ and\ \citenamefont {Urban}}]{blucher:2020}%
  \BibitemOpen
  \bibfield  {author} {\bibinfo {author} {\bibfnamefont {S.}~\bibnamefont
  {Bl\"ucher}}, \bibinfo {author} {\bibfnamefont {L.}~\bibnamefont {Kades}},
  \bibinfo {author} {\bibfnamefont {J.~M.}\ \bibnamefont {Pawlowski}}, \bibinfo
  {author} {\bibfnamefont {N.}~\bibnamefont {Strodthoff}}, \ and\ \bibinfo
  {author} {\bibfnamefont {J.~M.}\ \bibnamefont {Urban}},\ }\href {\doibase
  10.1103/PhysRevD.101.094507} {\bibfield  {journal} {\bibinfo  {journal}
  {Phys. Rev. D}\ }\textbf {\bibinfo {volume} {101}},\ \bibinfo {pages}
  {094507} (\bibinfo {year} {2020})}\BibitemShut {NoStop}%
\bibitem [{\citenamefont {Zhang}\ \emph {et~al.}(2020)\citenamefont {Zhang},
  \citenamefont {Ginsparg},\ and\ \citenamefont {Kim}}]{zhang:2020}%
  \BibitemOpen
  \bibfield  {author} {\bibinfo {author} {\bibfnamefont {Y.}~\bibnamefont
  {Zhang}}, \bibinfo {author} {\bibfnamefont {P.}~\bibnamefont {Ginsparg}}, \
  and\ \bibinfo {author} {\bibfnamefont {E.-A.}\ \bibnamefont {Kim}},\ }\href
  {\doibase 10.1103/PhysRevResearch.2.023283} {\bibfield  {journal} {\bibinfo
  {journal} {Phys. Rev. Research}\ }\textbf {\bibinfo {volume} {2}},\ \bibinfo
  {pages} {023283} (\bibinfo {year} {2020})}\BibitemShut {NoStop}%
\bibitem [{\citenamefont {Liu}\ \emph {et~al.}(2019)\citenamefont {Liu},
  \citenamefont {Greitemann},\ and\ \citenamefont {Pollet}}]{liu:2019}%
  \BibitemOpen
  \bibfield  {author} {\bibinfo {author} {\bibfnamefont {K.}~\bibnamefont
  {Liu}}, \bibinfo {author} {\bibfnamefont {J.}~\bibnamefont {Greitemann}}, \
  and\ \bibinfo {author} {\bibfnamefont {L.}~\bibnamefont {Pollet}},\ }\href
  {\doibase 10.1103/PhysRevB.99.104410} {\bibfield  {journal} {\bibinfo
  {journal} {Phys. Rev. B}\ }\textbf {\bibinfo {volume} {99}},\ \bibinfo
  {pages} {104410} (\bibinfo {year} {2019})}\BibitemShut {NoStop}%
\bibitem [{\citenamefont {Balabanov}\ and\ \citenamefont
  {Granath}(2020)}]{balabanov:2020}%
  \BibitemOpen
  \bibfield  {author} {\bibinfo {author} {\bibfnamefont {O.}~\bibnamefont
  {Balabanov}}\ and\ \bibinfo {author} {\bibfnamefont {M.}~\bibnamefont
  {Granath}},\ }\href {\doibase 10.1103/PhysRevResearch.2.013354} {\bibfield
  {journal} {\bibinfo  {journal} {Phys. Rev. Research}\ }\textbf {\bibinfo
  {volume} {2}},\ \bibinfo {pages} {013354} (\bibinfo {year}
  {2020})}\BibitemShut {NoStop}%
\bibitem [{\citenamefont {Long}\ \emph {et~al.}(2020)\citenamefont {Long},
  \citenamefont {Ren},\ and\ \citenamefont {Chen}}]{yang:2020}%
  \BibitemOpen
  \bibfield  {author} {\bibinfo {author} {\bibfnamefont {Y.}~\bibnamefont
  {Long}}, \bibinfo {author} {\bibfnamefont {J.}~\bibnamefont {Ren}}, \ and\
  \bibinfo {author} {\bibfnamefont {H.}~\bibnamefont {Chen}},\ }\href {\doibase
  10.1103/PhysRevLett.124.185501} {\bibfield  {journal} {\bibinfo  {journal}
  {Phys. Rev. Lett.}\ }\textbf {\bibinfo {volume} {124}},\ \bibinfo {pages}
  {185501} (\bibinfo {year} {2020})}\BibitemShut {NoStop}%
\bibitem [{\citenamefont {Singh}\ \emph {et~al.}(2020)\citenamefont {Singh},
  \citenamefont {Arora}, \citenamefont {Gupta},\ and\ \citenamefont
  {Scheurer}}]{singh:2020}%
  \BibitemOpen
  \bibfield  {author} {\bibinfo {author} {\bibfnamefont {J.}~\bibnamefont
  {Singh}}, \bibinfo {author} {\bibfnamefont {V.}~\bibnamefont {Arora}},
  \bibinfo {author} {\bibfnamefont {V.}~\bibnamefont {Gupta}}, \ and\ \bibinfo
  {author} {\bibfnamefont {M.~S.}\ \bibnamefont {Scheurer}},\ }\href
  {https://arxiv.org/abs/2006.11868} {\bibfield  {journal} {\bibinfo  {journal}
  {arXiv:2006.11868}\ } (\bibinfo {year} {2020})}\BibitemShut {NoStop}%
\bibitem [{\citenamefont {Bengio}\ and\ \citenamefont
  {Delalleau}(2011)}]{bengio:2011}%
  \BibitemOpen
  \bibfield  {author} {\bibinfo {author} {\bibfnamefont {Y.}~\bibnamefont
  {Bengio}}\ and\ \bibinfo {author} {\bibfnamefont {O.}~\bibnamefont
  {Delalleau}},\ }in\ \href {\doibase 10.1007/978-3-642-24412-4_3} {\emph
  {\bibinfo {booktitle} {Algorithmic Learning Theory}}},\ \bibinfo {editor}
  {edited by\ \bibinfo {editor} {\bibfnamefont {J.}~\bibnamefont {Kivinen}},
  \bibinfo {editor} {\bibfnamefont {C.}~\bibnamefont {Szepesv{\'a}ri}},
  \bibinfo {editor} {\bibfnamefont {E.}~\bibnamefont {Ukkonen}}, \ and\
  \bibinfo {editor} {\bibfnamefont {T.}~\bibnamefont {Zeugmann}}}\ (\bibinfo
  {publisher} {Springer Berlin Heidelberg},\ \bibinfo {address} {Berlin,
  Heidelberg},\ \bibinfo {year} {2011})\ pp.\ \bibinfo {pages}
  {18--36}\BibitemShut {NoStop}%
\bibitem [{\citenamefont {Goodfellow}\ \emph {et~al.}(2016)\citenamefont
  {Goodfellow}, \citenamefont {Bengio},\ and\ \citenamefont
  {Courville}}]{goodfellow:2016}%
  \BibitemOpen
  \bibfield  {author} {\bibinfo {author} {\bibfnamefont {I.}~\bibnamefont
  {Goodfellow}}, \bibinfo {author} {\bibfnamefont {Y.}~\bibnamefont {Bengio}},
  \ and\ \bibinfo {author} {\bibfnamefont {A.}~\bibnamefont {Courville}},\
  }\href {http://www.deeplearningbook.org} {\emph {\bibinfo {title} {Deep
  {L}earning}}}\ (\bibinfo  {publisher} {MIT Press},\ \bibinfo {year}
  {2016})\BibitemShut {NoStop}%
\bibitem [{\citenamefont {Molnar}(2019)}]{molnar:2019}%
  \BibitemOpen
  \bibfield  {author} {\bibinfo {author} {\bibfnamefont {C.}~\bibnamefont
  {Molnar}},\ }\href@noop {} {\emph {\bibinfo {title} {Interpretable Machine
  Learning}}}\ (\bibinfo {year} {2019})\ \bibinfo {note}
  {\url{https://christophm.github.io/interpretable-ml-book/}}\BibitemShut
  {NoStop}%
\bibitem [{\citenamefont {Cole}\ \emph {et~al.}(2020)\citenamefont {Cole},
  \citenamefont {Loges},\ and\ \citenamefont {Shiu}}]{cole:2020}%
  \BibitemOpen
  \bibfield  {author} {\bibinfo {author} {\bibfnamefont {A.}~\bibnamefont
  {Cole}}, \bibinfo {author} {\bibfnamefont {G.~J.}\ \bibnamefont {Loges}}, \
  and\ \bibinfo {author} {\bibfnamefont {G.}~\bibnamefont {Shiu}},\ }\href
  {https://arxiv.org/abs/2009.14231} {\bibfield  {journal} {\bibinfo  {journal}
  {arXiv:2009.14231}\ } (\bibinfo {year} {2020})}\BibitemShut {NoStop}%
\bibitem [{\citenamefont {Falicov}\ and\ \citenamefont
  {Kimball}(1969)}]{FKM:1969}%
  \BibitemOpen
  \bibfield  {author} {\bibinfo {author} {\bibfnamefont {L.~M.}\ \bibnamefont
  {Falicov}}\ and\ \bibinfo {author} {\bibfnamefont {J.~C.}\ \bibnamefont
  {Kimball}},\ }\href {\doibase 10.1103/PhysRevLett.22.997} {\bibfield
  {journal} {\bibinfo  {journal} {Phys. Rev. Lett.}\ }\textbf {\bibinfo
  {volume} {22}},\ \bibinfo {pages} {997} (\bibinfo {year} {1969})}\BibitemShut
  {NoStop}%
\bibitem [{\citenamefont {J.}(1963)}]{hubbard:1963}%
  \BibitemOpen
  \bibfield  {author} {\bibinfo {author} {\bibfnamefont {H.}~\bibnamefont
  {J.}},\ }\href {\doibase 10.1098/rspa.1963.0204} {\bibfield  {journal}
  {\bibinfo  {journal} {Proc. Royal Soc. Lond. A}\ }\textbf {\bibinfo {volume}
  {276}},\ \bibinfo {pages} {238} (\bibinfo {year} {1963})}\BibitemShut
  {NoStop}%
\bibitem [{\citenamefont {Freericks}\ and\ \citenamefont
  {Zlatic}(2003)}]{freericks:2003}%
  \BibitemOpen
  \bibfield  {author} {\bibinfo {author} {\bibfnamefont {J.~K.}\ \bibnamefont
  {Freericks}}\ and\ \bibinfo {author} {\bibfnamefont {V.}~\bibnamefont
  {Zlatic}},\ }\href {\doibase 10.1103/RevModPhys.75.1333} {\bibfield
  {journal} {\bibinfo  {journal} {Rev. Mod. Phys.}\ }\textbf {\bibinfo {volume}
  {75}},\ \bibinfo {pages} {1333} (\bibinfo {year} {2003})}\BibitemShut
  {NoStop}%
\bibitem [{\citenamefont {Lema\ifmmode~\acute{n}\else \'{n}\fi{}ski}\ \emph
  {et~al.}(2002)\citenamefont {Lema\ifmmode~\acute{n}\else \'{n}\fi{}ski},
  \citenamefont {Freericks},\ and\ \citenamefont {Banach}}]{lemanski:2002}%
  \BibitemOpen
  \bibfield  {author} {\bibinfo {author} {\bibfnamefont {R.}~\bibnamefont
  {Lema\ifmmode~\acute{n}\else \'{n}\fi{}ski}}, \bibinfo {author}
  {\bibfnamefont {J.~K.}\ \bibnamefont {Freericks}}, \ and\ \bibinfo {author}
  {\bibfnamefont {G.}~\bibnamefont {Banach}},\ }\href {\doibase
  10.1103/PhysRevLett.89.196403} {\bibfield  {journal} {\bibinfo  {journal}
  {Phys. Rev. Lett.}\ }\textbf {\bibinfo {volume} {89}},\ \bibinfo {pages}
  {196403} (\bibinfo {year} {2002})}\BibitemShut {NoStop}%
\bibitem [{\citenamefont {Lema{\'{n}}ski}\ \emph {et~al.}(2004)\citenamefont
  {Lema{\'{n}}ski}, \citenamefont {Freericks},\ and\ \citenamefont
  {Banach}}]{lemanski:2004}%
  \BibitemOpen
  \bibfield  {author} {\bibinfo {author} {\bibfnamefont {R.}~\bibnamefont
  {Lema{\'{n}}ski}}, \bibinfo {author} {\bibfnamefont {J.~K.}\ \bibnamefont
  {Freericks}}, \ and\ \bibinfo {author} {\bibfnamefont {G.}~\bibnamefont
  {Banach}},\ }\href {\doibase 10.1023/B:JOSS.0000037213.25834.33} {\bibfield
  {journal} {\bibinfo  {journal} {Journal of Statistical Physics}\ }\textbf
  {\bibinfo {volume} {116}},\ \bibinfo {pages} {699} (\bibinfo {year}
  {2004})}\BibitemShut {NoStop}%
\bibitem [{\citenamefont {Cencarikova}\ and\ \citenamefont
  {Farkasovsk{\`y}}(2011)}]{cencarikova:2012}%
  \BibitemOpen
  \bibfield  {author} {\bibinfo {author} {\bibfnamefont {H.}~\bibnamefont
  {Cencarikova}}\ and\ \bibinfo {author} {\bibfnamefont {P.}~\bibnamefont
  {Farkasovsk{\`y}}},\ }\href {\doibase 10.5488/CMP.14.42701} {\bibfield
  {journal} {\bibinfo  {journal} {Condens. Matter Phys.}\ }\textbf {\bibinfo
  {volume} {14}},\ \bibinfo {pages} {42701:1} (\bibinfo {year}
  {2011})}\BibitemShut {NoStop}%
\bibitem [{\citenamefont {Plischke}(1972)}]{plischke:1972}%
  \BibitemOpen
  \bibfield  {author} {\bibinfo {author} {\bibfnamefont {M.}~\bibnamefont
  {Plischke}},\ }\href {\doibase 10.1103/PhysRevLett.28.361} {\bibfield
  {journal} {\bibinfo  {journal} {Phys. Rev. Lett.}\ }\textbf {\bibinfo
  {volume} {28}},\ \bibinfo {pages} {361} (\bibinfo {year} {1972})}\BibitemShut
  {NoStop}%
\bibitem [{\citenamefont
  {Farka{\v{s}}ovsk{\'{y}}}(1995{\natexlab{a}})}]{farkasovsky:1995a}%
  \BibitemOpen
  \bibfield  {author} {\bibinfo {author} {\bibfnamefont {P.}~\bibnamefont
  {Farka{\v{s}}ovsk{\'{y}}}},\ }\href {\doibase 10.1103/PhysRevB.51.1507}
  {\bibfield  {journal} {\bibinfo  {journal} {Phys. Rev. B}\ }\textbf {\bibinfo
  {volume} {51}},\ \bibinfo {pages} {1507} (\bibinfo {year}
  {1995}{\natexlab{a}})}\BibitemShut {NoStop}%
\bibitem [{\citenamefont
  {Farka{\v{s}}ovsk{\'{y}}}(1995{\natexlab{b}})}]{farkasovsky:1995b}%
  \BibitemOpen
  \bibfield  {author} {\bibinfo {author} {\bibfnamefont {P.}~\bibnamefont
  {Farka{\v{s}}ovsk{\'{y}}}},\ }\href {\doibase 10.1103/PhysRevB.52.R5463}
  {\bibfield  {journal} {\bibinfo  {journal} {Phys. Rev. B}\ }\textbf {\bibinfo
  {volume} {52}},\ \bibinfo {pages} {R5463} (\bibinfo {year}
  {1995}{\natexlab{b}})}\BibitemShut {NoStop}%
\bibitem [{\citenamefont {Haldar}\ \emph {et~al.}(2019)\citenamefont {Haldar},
  \citenamefont {Laad},\ and\ \citenamefont {Hassan}}]{haldar:2019}%
  \BibitemOpen
  \bibfield  {author} {\bibinfo {author} {\bibfnamefont {P.}~\bibnamefont
  {Haldar}}, \bibinfo {author} {\bibfnamefont {M.~S.}\ \bibnamefont {Laad}}, \
  and\ \bibinfo {author} {\bibfnamefont {S.~R.}\ \bibnamefont {Hassan}},\
  }\href {\doibase 10.1103/PhysRevB.99.125147} {\bibfield  {journal} {\bibinfo
  {journal} {Phys. Rev. B}\ }\textbf {\bibinfo {volume} {99}},\ \bibinfo
  {pages} {125147} (\bibinfo {year} {2019})}\BibitemShut {NoStop}%
\bibitem [{\citenamefont {Kauch}\ \emph {et~al.}(2020)\citenamefont {Kauch},
  \citenamefont {Pudleiner}, \citenamefont {Astleithner}, \citenamefont
  {Thunstr\"om}, \citenamefont {Ribic},\ and\ \citenamefont
  {Held}}]{kauch:2020}%
  \BibitemOpen
  \bibfield  {author} {\bibinfo {author} {\bibfnamefont {A.}~\bibnamefont
  {Kauch}}, \bibinfo {author} {\bibfnamefont {P.}~\bibnamefont {Pudleiner}},
  \bibinfo {author} {\bibfnamefont {K.}~\bibnamefont {Astleithner}}, \bibinfo
  {author} {\bibfnamefont {P.}~\bibnamefont {Thunstr\"om}}, \bibinfo {author}
  {\bibfnamefont {T.}~\bibnamefont {Ribic}}, \ and\ \bibinfo {author}
  {\bibfnamefont {K.}~\bibnamefont {Held}},\ }\href {\doibase
  10.1103/PhysRevLett.124.047401} {\bibfield  {journal} {\bibinfo  {journal}
  {Phys. Rev. Lett.}\ }\textbf {\bibinfo {volume} {124}},\ \bibinfo {pages}
  {047401} (\bibinfo {year} {2020})}\BibitemShut {NoStop}%
\bibitem [{\citenamefont {Ma\ifmmode~\acute{s}\else \'{s}\fi{}ka}\ \emph
  {et~al.}(2008)\citenamefont {Ma\ifmmode~\acute{s}\else \'{s}\fi{}ka},
  \citenamefont {Lema\ifmmode~\acute{n}\else \'{n}\fi{}ski}, \citenamefont
  {Freericks},\ and\ \citenamefont {Williams}}]{maska:2008}%
  \BibitemOpen
  \bibfield  {author} {\bibinfo {author} {\bibfnamefont {M.~M.}\ \bibnamefont
  {Ma\ifmmode~\acute{s}\else \'{s}\fi{}ka}}, \bibinfo {author} {\bibfnamefont
  {R.}~\bibnamefont {Lema\ifmmode~\acute{n}\else \'{n}\fi{}ski}}, \bibinfo
  {author} {\bibfnamefont {J.~K.}\ \bibnamefont {Freericks}}, \ and\ \bibinfo
  {author} {\bibfnamefont {C.~J.}\ \bibnamefont {Williams}},\ }\href {\doibase
  10.1103/PhysRevLett.101.060404} {\bibfield  {journal} {\bibinfo  {journal}
  {Phys. Rev. Lett.}\ }\textbf {\bibinfo {volume} {101}},\ \bibinfo {pages}
  {060404} (\bibinfo {year} {2008})}\BibitemShut {NoStop}%
\bibitem [{\citenamefont {Ma\ifmmode~\acute{s}\else \'{s}\fi{}ka}\ \emph
  {et~al.}(2011)\citenamefont {Ma\ifmmode~\acute{s}\else \'{s}\fi{}ka},
  \citenamefont {Lema\ifmmode~\acute{n}\else \'{n}\fi{}ski}, \citenamefont
  {Williams},\ and\ \citenamefont {Freericks}}]{maska:2011}%
  \BibitemOpen
  \bibfield  {author} {\bibinfo {author} {\bibfnamefont {M.~M.}\ \bibnamefont
  {Ma\ifmmode~\acute{s}\else \'{s}\fi{}ka}}, \bibinfo {author} {\bibfnamefont
  {R.}~\bibnamefont {Lema\ifmmode~\acute{n}\else \'{n}\fi{}ski}}, \bibinfo
  {author} {\bibfnamefont {C.~J.}\ \bibnamefont {Williams}}, \ and\ \bibinfo
  {author} {\bibfnamefont {J.~K.}\ \bibnamefont {Freericks}},\ }\href {\doibase
  10.1103/PhysRevA.83.063631} {\bibfield  {journal} {\bibinfo  {journal} {Phys.
  Rev. A}\ }\textbf {\bibinfo {volume} {83}},\ \bibinfo {pages} {063631}
  (\bibinfo {year} {2011})}\BibitemShut {NoStop}%
\bibitem [{\citenamefont {Hu}\ \emph {et~al.}(2015)\citenamefont {Hu},
  \citenamefont {Ma\ifmmode~\acute{s}\else \'{s}\fi{}ka}, \citenamefont
  {Clark},\ and\ \citenamefont {Freericks}}]{hu:2015}%
  \BibitemOpen
  \bibfield  {author} {\bibinfo {author} {\bibfnamefont {A.}~\bibnamefont
  {Hu}}, \bibinfo {author} {\bibfnamefont {M.~M.}\ \bibnamefont
  {Ma\ifmmode~\acute{s}\else \'{s}\fi{}ka}}, \bibinfo {author} {\bibfnamefont
  {C.~W.}\ \bibnamefont {Clark}}, \ and\ \bibinfo {author} {\bibfnamefont
  {J.~K.}\ \bibnamefont {Freericks}},\ }\href {\doibase
  10.1103/PhysRevA.91.063624} {\bibfield  {journal} {\bibinfo  {journal} {Phys.
  Rev. A}\ }\textbf {\bibinfo {volume} {91}},\ \bibinfo {pages} {063624}
  (\bibinfo {year} {2015})}\BibitemShut {NoStop}%
\bibitem [{\citenamefont {Qin}\ \emph {et~al.}(2018)\citenamefont {Qin},
  \citenamefont {Schnell}, \citenamefont {Sengstock}, \citenamefont
  {Weitenberg}, \citenamefont {Eckardt},\ and\ \citenamefont
  {Hofstetter}}]{qin:2018}%
  \BibitemOpen
  \bibfield  {author} {\bibinfo {author} {\bibfnamefont {T.}~\bibnamefont
  {Qin}}, \bibinfo {author} {\bibfnamefont {A.}~\bibnamefont {Schnell}},
  \bibinfo {author} {\bibfnamefont {K.}~\bibnamefont {Sengstock}}, \bibinfo
  {author} {\bibfnamefont {C.}~\bibnamefont {Weitenberg}}, \bibinfo {author}
  {\bibfnamefont {A.}~\bibnamefont {Eckardt}}, \ and\ \bibinfo {author}
  {\bibfnamefont {W.}~\bibnamefont {Hofstetter}},\ }\href {\doibase
  10.1103/PhysRevA.98.033601} {\bibfield  {journal} {\bibinfo  {journal} {Phys.
  Rev. A}\ }\textbf {\bibinfo {volume} {98}},\ \bibinfo {pages} {033601}
  (\bibinfo {year} {2018})}\BibitemShut {NoStop}%
\bibitem [{\citenamefont {Maionchi}\ \emph {et~al.}(2008)\citenamefont
  {Maionchi}, \citenamefont {Souza}, \citenamefont {Herrmann},\ and\
  \citenamefont {da~Costa~Filho}}]{maionchi:2008}%
  \BibitemOpen
  \bibfield  {author} {\bibinfo {author} {\bibfnamefont {D.~O.}\ \bibnamefont
  {Maionchi}}, \bibinfo {author} {\bibfnamefont {A.~M.~C.}\ \bibnamefont
  {Souza}}, \bibinfo {author} {\bibfnamefont {H.~J.}\ \bibnamefont {Herrmann}},
  \ and\ \bibinfo {author} {\bibfnamefont {R.~N.}\ \bibnamefont
  {da~Costa~Filho}},\ }\href {\doibase 10.1103/PhysRevB.77.245126} {\bibfield
  {journal} {\bibinfo  {journal} {Phys. Rev. B}\ }\textbf {\bibinfo {volume}
  {77}},\ \bibinfo {pages} {245126} (\bibinfo {year} {2008})}\BibitemShut
  {NoStop}%
\bibitem [{\citenamefont {Antipov}\ \emph {et~al.}(2016)\citenamefont
  {Antipov}, \citenamefont {Javanmard}, \citenamefont {Ribeiro},\ and\
  \citenamefont {Kirchner}}]{antipov:2016}%
  \BibitemOpen
  \bibfield  {author} {\bibinfo {author} {\bibfnamefont {A.~E.}\ \bibnamefont
  {Antipov}}, \bibinfo {author} {\bibfnamefont {Y.}~\bibnamefont {Javanmard}},
  \bibinfo {author} {\bibfnamefont {P.}~\bibnamefont {Ribeiro}}, \ and\
  \bibinfo {author} {\bibfnamefont {S.}~\bibnamefont {Kirchner}},\ }\href
  {\doibase 10.1103/PhysRevLett.117.146601} {\bibfield  {journal} {\bibinfo
  {journal} {Phys. Rev. Lett.}\ }\textbf {\bibinfo {volume} {117}},\ \bibinfo
  {pages} {146601} (\bibinfo {year} {2016})}\BibitemShut {NoStop}%
\bibitem [{\citenamefont {Haldar}\ \emph {et~al.}(2017)\citenamefont {Haldar},
  \citenamefont {Laad},\ and\ \citenamefont {Hassan}}]{haldar:2017b}%
  \BibitemOpen
  \bibfield  {author} {\bibinfo {author} {\bibfnamefont {P.}~\bibnamefont
  {Haldar}}, \bibinfo {author} {\bibfnamefont {M.~S.}\ \bibnamefont {Laad}}, \
  and\ \bibinfo {author} {\bibfnamefont {S.~R.}\ \bibnamefont {Hassan}},\
  }\href {\doibase 10.1103/PhysRevB.95.125116} {\bibfield  {journal} {\bibinfo
  {journal} {Phys. Rev. B}\ }\textbf {\bibinfo {volume} {95}},\ \bibinfo
  {pages} {125116} (\bibinfo {year} {2017})}\BibitemShut {NoStop}%
\bibitem [{\citenamefont {Smith}\ \emph {et~al.}(2017)\citenamefont {Smith},
  \citenamefont {Knolle}, \citenamefont {Kovrizhin},\ and\ \citenamefont
  {Moessner}}]{smith:2017}%
  \BibitemOpen
  \bibfield  {author} {\bibinfo {author} {\bibfnamefont {A.}~\bibnamefont
  {Smith}}, \bibinfo {author} {\bibfnamefont {J.}~\bibnamefont {Knolle}},
  \bibinfo {author} {\bibfnamefont {D.~L.}\ \bibnamefont {Kovrizhin}}, \ and\
  \bibinfo {author} {\bibfnamefont {R.}~\bibnamefont {Moessner}},\ }\href
  {\doibase 10.1103/PhysRevLett.118.266601} {\bibfield  {journal} {\bibinfo
  {journal} {Phys. Rev. Lett.}\ }\textbf {\bibinfo {volume} {118}},\ \bibinfo
  {pages} {266601} (\bibinfo {year} {2017})}\BibitemShut {NoStop}%
\bibitem [{\citenamefont {\ifmmode~\check{Z}\else \v{Z}\fi{}onda}\ \emph
  {et~al.}(2019)\citenamefont {\ifmmode~\check{Z}\else \v{Z}\fi{}onda},
  \citenamefont {Okamoto},\ and\ \citenamefont {Thoss}}]{zonda:2019b}%
  \BibitemOpen
  \bibfield  {author} {\bibinfo {author} {\bibfnamefont {M.}~\bibnamefont
  {\ifmmode~\check{Z}\else \v{Z}\fi{}onda}}, \bibinfo {author} {\bibfnamefont
  {J.}~\bibnamefont {Okamoto}}, \ and\ \bibinfo {author} {\bibfnamefont
  {M.}~\bibnamefont {Thoss}},\ }\href {\doibase 10.1103/PhysRevB.100.075124}
  {\bibfield  {journal} {\bibinfo  {journal} {Phys. Rev. B}\ }\textbf {\bibinfo
  {volume} {100}},\ \bibinfo {pages} {075124} (\bibinfo {year}
  {2019})}\BibitemShut {NoStop}%
\bibitem [{\citenamefont {Ribic}\ \emph {et~al.}(2016)\citenamefont {Ribic},
  \citenamefont {Rohringer},\ and\ \citenamefont {Held}}]{ribic:2016}%
  \BibitemOpen
  \bibfield  {author} {\bibinfo {author} {\bibfnamefont {T.}~\bibnamefont
  {Ribic}}, \bibinfo {author} {\bibfnamefont {G.}~\bibnamefont {Rohringer}}, \
  and\ \bibinfo {author} {\bibfnamefont {K.}~\bibnamefont {Held}},\ }\href
  {\doibase 10.1103/PhysRevB.93.195105} {\bibfield  {journal} {\bibinfo
  {journal} {Phys. Rev. B}\ }\textbf {\bibinfo {volume} {93}},\ \bibinfo
  {pages} {195105} (\bibinfo {year} {2016})}\BibitemShut {NoStop}%
\bibitem [{\citenamefont {Ribic}\ \emph {et~al.}(2017)\citenamefont {Ribic},
  \citenamefont {Rohringer},\ and\ \citenamefont {Held}}]{ribic:2017}%
  \BibitemOpen
  \bibfield  {author} {\bibinfo {author} {\bibfnamefont {T.}~\bibnamefont
  {Ribic}}, \bibinfo {author} {\bibfnamefont {G.}~\bibnamefont {Rohringer}}, \
  and\ \bibinfo {author} {\bibfnamefont {K.}~\bibnamefont {Held}},\ }\href
  {\doibase 10.1103/PhysRevB.95.155130} {\bibfield  {journal} {\bibinfo
  {journal} {Phys. Rev. B}\ }\textbf {\bibinfo {volume} {95}},\ \bibinfo
  {pages} {155130} (\bibinfo {year} {2017})}\BibitemShut {NoStop}%
\bibitem [{\citenamefont {Freericks}\ \emph {et~al.}(2006)\citenamefont
  {Freericks}, \citenamefont {Turkowski},\ and\ \citenamefont
  {Zlati\ifmmode~\acute{c}\else \'{c}\fi{}}}]{freericks:2006}%
  \BibitemOpen
  \bibfield  {author} {\bibinfo {author} {\bibfnamefont {J.~K.}\ \bibnamefont
  {Freericks}}, \bibinfo {author} {\bibfnamefont {V.~M.}\ \bibnamefont
  {Turkowski}}, \ and\ \bibinfo {author} {\bibfnamefont {V.}~\bibnamefont
  {Zlati\ifmmode~\acute{c}\else \'{c}\fi{}}},\ }\href {\doibase
  10.1103/PhysRevLett.97.266408} {\bibfield  {journal} {\bibinfo  {journal}
  {Phys. Rev. Lett.}\ }\textbf {\bibinfo {volume} {97}},\ \bibinfo {pages}
  {266408} (\bibinfo {year} {2006})}\BibitemShut {NoStop}%
\bibitem [{\citenamefont {Eckstein}\ and\ \citenamefont
  {Kollar}(2008)}]{eckstein:2008}%
  \BibitemOpen
  \bibfield  {author} {\bibinfo {author} {\bibfnamefont {M.}~\bibnamefont
  {Eckstein}}\ and\ \bibinfo {author} {\bibfnamefont {M.}~\bibnamefont
  {Kollar}},\ }\href {\doibase 10.1103/PhysRevLett.100.120404} {\bibfield
  {journal} {\bibinfo  {journal} {Phys. Rev. Lett.}\ }\textbf {\bibinfo
  {volume} {100}},\ \bibinfo {pages} {120404} (\bibinfo {year}
  {2008})}\BibitemShut {NoStop}%
\bibitem [{\citenamefont {Eckstein}\ \emph {et~al.}(2009)\citenamefont
  {Eckstein}, \citenamefont {Kollar},\ and\ \citenamefont
  {Werner}}]{eckstein:2009}%
  \BibitemOpen
  \bibfield  {author} {\bibinfo {author} {\bibfnamefont {M.}~\bibnamefont
  {Eckstein}}, \bibinfo {author} {\bibfnamefont {M.}~\bibnamefont {Kollar}}, \
  and\ \bibinfo {author} {\bibfnamefont {P.}~\bibnamefont {Werner}},\ }\href
  {\doibase 10.1103/PhysRevLett.103.056403} {\bibfield  {journal} {\bibinfo
  {journal} {Phys. Rev. Lett.}\ }\textbf {\bibinfo {volume} {103}},\ \bibinfo
  {pages} {056403} (\bibinfo {year} {2009})}\BibitemShut {NoStop}%
\bibitem [{\citenamefont {Herrmann}\ \emph {et~al.}(2016)\citenamefont
  {Herrmann}, \citenamefont {Tsuji}, \citenamefont {Eckstein},\ and\
  \citenamefont {Werner}}]{herrmann:2016}%
  \BibitemOpen
  \bibfield  {author} {\bibinfo {author} {\bibfnamefont {A.~J.}\ \bibnamefont
  {Herrmann}}, \bibinfo {author} {\bibfnamefont {N.}~\bibnamefont {Tsuji}},
  \bibinfo {author} {\bibfnamefont {M.}~\bibnamefont {Eckstein}}, \ and\
  \bibinfo {author} {\bibfnamefont {P.}~\bibnamefont {Werner}},\ }\href
  {\doibase 10.1103/PhysRevB.94.245114} {\bibfield  {journal} {\bibinfo
  {journal} {Phys. Rev. B}\ }\textbf {\bibinfo {volume} {94}},\ \bibinfo
  {pages} {245114} (\bibinfo {year} {2016})}\BibitemShut {NoStop}%
\bibitem [{\citenamefont {Herrmann}\ \emph {et~al.}(2018)\citenamefont
  {Herrmann}, \citenamefont {Antipov},\ and\ \citenamefont
  {Werner}}]{herrmann:2018}%
  \BibitemOpen
  \bibfield  {author} {\bibinfo {author} {\bibfnamefont {A.~J.}\ \bibnamefont
  {Herrmann}}, \bibinfo {author} {\bibfnamefont {A.~E.}\ \bibnamefont
  {Antipov}}, \ and\ \bibinfo {author} {\bibfnamefont {P.}~\bibnamefont
  {Werner}},\ }\href {\doibase 10.1103/PhysRevB.97.165107} {\bibfield
  {journal} {\bibinfo  {journal} {Phys. Rev. B}\ }\textbf {\bibinfo {volume}
  {97}},\ \bibinfo {pages} {165107} (\bibinfo {year} {2018})}\BibitemShut
  {NoStop}%
\bibitem [{\citenamefont {Freericks}(2006)}]{FreericksBook2006}%
  \BibitemOpen
  \bibfield  {author} {\bibinfo {author} {\bibfnamefont {J.}~\bibnamefont
  {Freericks}},\ }\href {\doibase doi:10.1142/p1053} {\emph {\bibinfo {title}
  {Transport in Multilayered Nanostructures: The Dynamical Mean-Field Theory
  Approach}}}\ (\bibinfo  {publisher} {Imperial College Press, London},\
  \bibinfo {year} {2006})\ pp.\ \bibinfo {pages} {1--328}\BibitemShut {NoStop}%
\bibitem [{\citenamefont {\ifmmode~\check{Z}\else \v{Z}\fi{}onda}\ and\
  \citenamefont {Thoss}(2019)}]{zonda:2019a}%
  \BibitemOpen
  \bibfield  {author} {\bibinfo {author} {\bibfnamefont {M.}~\bibnamefont
  {\ifmmode~\check{Z}\else \v{Z}\fi{}onda}}\ and\ \bibinfo {author}
  {\bibfnamefont {M.}~\bibnamefont {Thoss}},\ }\href {\doibase
  10.1103/PhysRevB.99.155157} {\bibfield  {journal} {\bibinfo  {journal} {Phys.
  Rev. B}\ }\textbf {\bibinfo {volume} {99}},\ \bibinfo {pages} {155157}
  (\bibinfo {year} {2019})}\BibitemShut {NoStop}%
\bibitem [{\citenamefont {Smorka}\ \emph {et~al.}(2020)\citenamefont {Smorka},
  \citenamefont {{\v{Z}}onda},\ and\ \citenamefont {Thoss}}]{smorka:2020}%
  \BibitemOpen
  \bibfield  {author} {\bibinfo {author} {\bibfnamefont {R.}~\bibnamefont
  {Smorka}}, \bibinfo {author} {\bibfnamefont {M.}~\bibnamefont {{\v{Z}}onda}},
  \ and\ \bibinfo {author} {\bibfnamefont {M.}~\bibnamefont {Thoss}},\
  }\href@noop {} {\bibfield  {journal} {\bibinfo  {journal} {Phys. Rev. B}\
  }\textbf {\bibinfo {volume} {101}},\ \bibinfo {pages} {155116} (\bibinfo
  {year} {2020})}\BibitemShut {NoStop}%
\bibitem [{\citenamefont {Schattschneider}(1978)}]{doris:1978}%
  \BibitemOpen
  \bibfield  {author} {\bibinfo {author} {\bibfnamefont {D.}~\bibnamefont
  {Schattschneider}},\ }\href {\doibase 10.1080/00029890.1978.11994612}
  {\bibfield  {journal} {\bibinfo  {journal} {The American Mathematical
  Monthly}\ }\textbf {\bibinfo {volume} {85}},\ \bibinfo {pages} {439}
  (\bibinfo {year} {1978})}\BibitemShut {NoStop}%
\bibitem [{Note1()}]{Note1}%
  \BibitemOpen
  \bibinfo {note} {See Supplemental Material for details on the simulated
  annealing procedure, the prediction-based, and the mean-based method, as well
  as a discussion of alternative phase classification methods and definitions
  of order parameters and correlations functions which includes Refs.~\cite
  {github,lemanski:2002,lemanski:2004,maska:2006,tran:2006,zonda:2012,goodfellow:2016,lecun:2012,paszke:2019,kingma:2014,blucher:2020,casert:2019,
  zhang:2020,scikit:2011,wetzel:2017,kreher:1999,chau:1998,wang:2016,mehta:2019,vaart:1998}.}\BibitemShut
  {Stop}%
\bibitem [{\citenamefont {Petrovi\ifmmode~\acute{c}\else \'{c}\fi{}}\ \emph
  {et~al.}(2018)\citenamefont {Petrovi\ifmmode~\acute{c}\else \'{c}\fi{}},
  \citenamefont {Popescu}, \citenamefont {Bajpai}, \citenamefont
  {Plech\'a\ifmmode~\check{c}\else \v{c}\fi{}},\ and\ \citenamefont
  {Nikoli\ifmmode~\acute{c}\else \'{c}\fi{}}}]{petrovic:2018}%
  \BibitemOpen
  \bibfield  {author} {\bibinfo {author} {\bibfnamefont {M.~D.}\ \bibnamefont
  {Petrovi\ifmmode~\acute{c}\else \'{c}\fi{}}}, \bibinfo {author}
  {\bibfnamefont {B.~S.}\ \bibnamefont {Popescu}}, \bibinfo {author}
  {\bibfnamefont {U.}~\bibnamefont {Bajpai}}, \bibinfo {author} {\bibfnamefont
  {P.}~\bibnamefont {Plech\'a\ifmmode~\check{c}\else \v{c}\fi{}}}, \ and\
  \bibinfo {author} {\bibfnamefont {B.~K.}\ \bibnamefont
  {Nikoli\ifmmode~\acute{c}\else \'{c}\fi{}}},\ }\href {\doibase
  10.1103/PhysRevApplied.10.054038} {\bibfield  {journal} {\bibinfo  {journal}
  {Phys. Rev. Applied}\ }\textbf {\bibinfo {volume} {10}},\ \bibinfo {pages}
  {054038} (\bibinfo {year} {2018})}\BibitemShut {NoStop}%
\bibitem [{\citenamefont {Li}\ \emph {et~al.}(2019)\citenamefont {Li},
  \citenamefont {Chen},\ and\ \citenamefont {Ng}}]{li:2019}%
  \BibitemOpen
  \bibfield  {author} {\bibinfo {author} {\bibfnamefont {X.-H.}\ \bibnamefont
  {Li}}, \bibinfo {author} {\bibfnamefont {Z.}~\bibnamefont {Chen}}, \ and\
  \bibinfo {author} {\bibfnamefont {T.~K.}\ \bibnamefont {Ng}},\ }\href
  {\doibase 10.1103/PhysRevB.100.094519} {\bibfield  {journal} {\bibinfo
  {journal} {Phys. Rev. B}\ }\textbf {\bibinfo {volume} {100}},\ \bibinfo
  {pages} {094519} (\bibinfo {year} {2019})}\BibitemShut {NoStop}%
\bibitem [{\citenamefont {Gon\ifmmode~\mbox{\c{c}}\else \c{c}\fi{}alves}\ \emph
  {et~al.}(2019)\citenamefont {Gon\ifmmode~\mbox{\c{c}}\else \c{c}\fi{}alves},
  \citenamefont {Ribeiro}, \citenamefont {Mondaini},\ and\ \citenamefont
  {Castro}}]{goncalves:2019}%
  \BibitemOpen
  \bibfield  {author} {\bibinfo {author} {\bibfnamefont {M.}~\bibnamefont
  {Gon\ifmmode~\mbox{\c{c}}\else \c{c}\fi{}alves}}, \bibinfo {author}
  {\bibfnamefont {P.}~\bibnamefont {Ribeiro}}, \bibinfo {author} {\bibfnamefont
  {R.}~\bibnamefont {Mondaini}}, \ and\ \bibinfo {author} {\bibfnamefont
  {E.~V.}\ \bibnamefont {Castro}},\ }\href {\doibase
  10.1103/PhysRevLett.122.126601} {\bibfield  {journal} {\bibinfo  {journal}
  {Phys. Rev. Lett.}\ }\textbf {\bibinfo {volume} {122}},\ \bibinfo {pages}
  {126601} (\bibinfo {year} {2019})}\BibitemShut {NoStop}%
\bibitem [{\citenamefont {Arnold}\ \emph {et~al.}(2020)\citenamefont {Arnold},
  \citenamefont {Sch\"{a}fer}, \citenamefont {{\v{Z}}onda},\ and\ \citenamefont
  {Lode}}]{github}%
  \BibitemOpen
  \bibfield  {author} {\bibinfo {author} {\bibfnamefont {J.}~\bibnamefont
  {Arnold}}, \bibinfo {author} {\bibfnamefont {F.}~\bibnamefont {Sch\"{a}fer}},
  \bibinfo {author} {\bibfnamefont {M.}~\bibnamefont {{\v{Z}}onda}}, \ and\
  \bibinfo {author} {\bibfnamefont {A.~U.~J.}\ \bibnamefont {Lode}},\
  }\href@noop {} {\emph {\bibinfo {title} {Interpretable and unsupervised phase
  classification}}}\ (\bibinfo {year} {2020})\ \bibinfo {note}
  {\url{https://github.com/arnoldjulian/Interpretable-and-unsupervised-phase-classification}}\BibitemShut
  {NoStop}%
\bibitem [{\citenamefont {Ma\ifmmode~\acute{s}\else \'{s}\fi{}ka}\ and\
  \citenamefont {Czajka}(2006)}]{maska:2006}%
  \BibitemOpen
  \bibfield  {author} {\bibinfo {author} {\bibfnamefont {M.~M.}\ \bibnamefont
  {Ma\ifmmode~\acute{s}\else \'{s}\fi{}ka}}\ and\ \bibinfo {author}
  {\bibfnamefont {K.}~\bibnamefont {Czajka}},\ }\href {\doibase
  10.1103/PhysRevB.74.035109} {\bibfield  {journal} {\bibinfo  {journal} {Phys.
  Rev. B}\ }\textbf {\bibinfo {volume} {74}},\ \bibinfo {pages} {035109}
  (\bibinfo {year} {2006})}\BibitemShut {NoStop}%
\bibitem [{\citenamefont {Tran}(2006)}]{tran:2006}%
  \BibitemOpen
  \bibfield  {author} {\bibinfo {author} {\bibfnamefont {M.-T.}\ \bibnamefont
  {Tran}},\ }\href {\doibase 10.1103/PhysRevB.73.205110} {\bibfield  {journal}
  {\bibinfo  {journal} {Phys. Rev. B}\ }\textbf {\bibinfo {volume} {73}},\
  \bibinfo {pages} {205110} (\bibinfo {year} {2006})}\BibitemShut {NoStop}%
\bibitem [{\citenamefont {Žonda}(2012)}]{zonda:2012}%
  \BibitemOpen
  \bibfield  {author} {\bibinfo {author} {\bibfnamefont {M.}~\bibnamefont
  {Žonda}},\ }\href {\doibase 10.1080/01411594.2011.604509} {\bibfield
  {journal} {\bibinfo  {journal} {Phase Transitions}\ }\textbf {\bibinfo
  {volume} {85}},\ \bibinfo {pages} {96} (\bibinfo {year} {2012})}\BibitemShut
  {NoStop}%
\bibitem [{\citenamefont {LeCun}\ \emph {et~al.}(2012)\citenamefont {LeCun},
  \citenamefont {Bottou}, \citenamefont {Orr},\ and\ \citenamefont
  {M{\"u}ller}}]{lecun:2012}%
  \BibitemOpen
  \bibfield  {author} {\bibinfo {author} {\bibfnamefont {Y.~A.}\ \bibnamefont
  {LeCun}}, \bibinfo {author} {\bibfnamefont {L.}~\bibnamefont {Bottou}},
  \bibinfo {author} {\bibfnamefont {G.~B.}\ \bibnamefont {Orr}}, \ and\
  \bibinfo {author} {\bibfnamefont {K.-R.}\ \bibnamefont {M{\"u}ller}},\ }in\
  \href {\doibase https://doi.org/10.1007/978-3-642-35289-8_3} {\emph {\bibinfo
  {booktitle} {Neural {N}etworks: {T}ricks of the {T}rade}}}\ (\bibinfo
  {publisher} {Springer},\ \bibinfo {year} {2012})\ pp.\ \bibinfo {pages}
  {9--48}\BibitemShut {NoStop}%
\bibitem [{\citenamefont {Paszke}\ \emph {et~al.}(2019)\citenamefont {Paszke},
  \citenamefont {Gross}, \citenamefont {Massa}, \citenamefont {Lerer},
  \citenamefont {Bradbury}, \citenamefont {Chanan}, \citenamefont {Killeen},
  \citenamefont {Lin}, \citenamefont {Gimelshein}, \citenamefont {Antiga},
  \citenamefont {Desmaison}, \citenamefont {Kopf}, \citenamefont {Yang},
  \citenamefont {DeVito}, \citenamefont {Raison}, \citenamefont {Tejani},
  \citenamefont {Chilamkurthy}, \citenamefont {Steiner}, \citenamefont {Fang},
  \citenamefont {Bai},\ and\ \citenamefont {Chintala}}]{paszke:2019}%
  \BibitemOpen
  \bibfield  {author} {\bibinfo {author} {\bibfnamefont {A.}~\bibnamefont
  {Paszke}}, \bibinfo {author} {\bibfnamefont {S.}~\bibnamefont {Gross}},
  \bibinfo {author} {\bibfnamefont {F.}~\bibnamefont {Massa}}, \bibinfo
  {author} {\bibfnamefont {A.}~\bibnamefont {Lerer}}, \bibinfo {author}
  {\bibfnamefont {J.}~\bibnamefont {Bradbury}}, \bibinfo {author}
  {\bibfnamefont {G.}~\bibnamefont {Chanan}}, \bibinfo {author} {\bibfnamefont
  {T.}~\bibnamefont {Killeen}}, \bibinfo {author} {\bibfnamefont
  {Z.}~\bibnamefont {Lin}}, \bibinfo {author} {\bibfnamefont {N.}~\bibnamefont
  {Gimelshein}}, \bibinfo {author} {\bibfnamefont {L.}~\bibnamefont {Antiga}},
  \bibinfo {author} {\bibfnamefont {A.}~\bibnamefont {Desmaison}}, \bibinfo
  {author} {\bibfnamefont {A.}~\bibnamefont {Kopf}}, \bibinfo {author}
  {\bibfnamefont {E.}~\bibnamefont {Yang}}, \bibinfo {author} {\bibfnamefont
  {Z.}~\bibnamefont {DeVito}}, \bibinfo {author} {\bibfnamefont
  {M.}~\bibnamefont {Raison}}, \bibinfo {author} {\bibfnamefont
  {A.}~\bibnamefont {Tejani}}, \bibinfo {author} {\bibfnamefont
  {S.}~\bibnamefont {Chilamkurthy}}, \bibinfo {author} {\bibfnamefont
  {B.}~\bibnamefont {Steiner}}, \bibinfo {author} {\bibfnamefont
  {L.}~\bibnamefont {Fang}}, \bibinfo {author} {\bibfnamefont {J.}~\bibnamefont
  {Bai}}, \ and\ \bibinfo {author} {\bibfnamefont {S.}~\bibnamefont
  {Chintala}},\ }in\ \href
  {http://papers.nips.cc/paper/9015-pytorch-an-imperative-style-high-performance-deep-learning-library.pdf}
  {\emph {\bibinfo {booktitle} {Advances in Neural Information Processing
  Systems 32}}},\ \bibinfo {editor} {edited by\ \bibinfo {editor}
  {\bibfnamefont {H.}~\bibnamefont {Wallach}}, \bibinfo {editor} {\bibfnamefont
  {H.}~\bibnamefont {Larochelle}}, \bibinfo {editor} {\bibfnamefont
  {A.}~\bibnamefont {Beygelzimer}}, \bibinfo {editor} {\bibfnamefont
  {F.}~\bibnamefont {d'Alch\'{e} Buc}}, \bibinfo {editor} {\bibfnamefont
  {E.}~\bibnamefont {Fox}}, \ and\ \bibinfo {editor} {\bibfnamefont
  {R.}~\bibnamefont {Garnett}}}\ (\bibinfo  {publisher} {Curran Associates,
  Inc.},\ \bibinfo {year} {2019})\ pp.\ \bibinfo {pages}
  {8026--8037}\BibitemShut {NoStop}%
\bibitem [{\citenamefont {Kingma}\ and\ \citenamefont
  {Ba}(2014)}]{kingma:2014}%
  \BibitemOpen
  \bibfield  {author} {\bibinfo {author} {\bibfnamefont {D.}~\bibnamefont
  {Kingma}}\ and\ \bibinfo {author} {\bibfnamefont {J.}~\bibnamefont {Ba}},\
  }\href {https://arxiv.org/abs/1412.6980} {\bibfield  {journal} {\bibinfo
  {journal} {arXiv:1412.6980}\ } (\bibinfo {year} {2014})}\BibitemShut
  {NoStop}%
\bibitem [{\citenamefont {Pedregosa}\ \emph {et~al.}(2011)\citenamefont
  {Pedregosa}, \citenamefont {Varoquaux}, \citenamefont {Gramfort},
  \citenamefont {Michel}, \citenamefont {Thirion}, \citenamefont {Grisel},
  \citenamefont {Blondel}, \citenamefont {Prettenhofer}, \citenamefont {Weiss},
  \citenamefont {Dubourg}, \citenamefont {Vanderplas}, \citenamefont {Passos},
  \citenamefont {Cournapeau}, \citenamefont {Brucher}, \citenamefont {Perrot},\
  and\ \citenamefont {{{\'E}}douard Duchesnay}}]{scikit:2011}%
  \BibitemOpen
  \bibfield  {author} {\bibinfo {author} {\bibfnamefont {F.}~\bibnamefont
  {Pedregosa}}, \bibinfo {author} {\bibfnamefont {G.}~\bibnamefont
  {Varoquaux}}, \bibinfo {author} {\bibfnamefont {A.}~\bibnamefont {Gramfort}},
  \bibinfo {author} {\bibfnamefont {V.}~\bibnamefont {Michel}}, \bibinfo
  {author} {\bibfnamefont {B.}~\bibnamefont {Thirion}}, \bibinfo {author}
  {\bibfnamefont {O.}~\bibnamefont {Grisel}}, \bibinfo {author} {\bibfnamefont
  {M.}~\bibnamefont {Blondel}}, \bibinfo {author} {\bibfnamefont
  {P.}~\bibnamefont {Prettenhofer}}, \bibinfo {author} {\bibfnamefont
  {R.}~\bibnamefont {Weiss}}, \bibinfo {author} {\bibfnamefont
  {V.}~\bibnamefont {Dubourg}}, \bibinfo {author} {\bibfnamefont
  {J.}~\bibnamefont {Vanderplas}}, \bibinfo {author} {\bibfnamefont
  {A.}~\bibnamefont {Passos}}, \bibinfo {author} {\bibfnamefont
  {D.}~\bibnamefont {Cournapeau}}, \bibinfo {author} {\bibfnamefont
  {M.}~\bibnamefont {Brucher}}, \bibinfo {author} {\bibfnamefont
  {M.}~\bibnamefont {Perrot}}, \ and\ \bibinfo {author} {\bibnamefont
  {{{\'E}}douard Duchesnay}},\ }\href
  {http://jmlr.org/papers/v12/pedregosa11a.html} {\bibfield  {journal}
  {\bibinfo  {journal} {Journal of Machine Learning Research}\ }\textbf
  {\bibinfo {volume} {12}},\ \bibinfo {pages} {2825} (\bibinfo {year}
  {2011})}\BibitemShut {NoStop}%
\bibitem [{\citenamefont {Kreher}\ and\ \citenamefont
  {Stinson}(1999)}]{kreher:1999}%
  \BibitemOpen
  \bibfield  {author} {\bibinfo {author} {\bibfnamefont {D.~L.}\ \bibnamefont
  {Kreher}}\ and\ \bibinfo {author} {\bibfnamefont {D.~R.}\ \bibnamefont
  {Stinson}},\ }\href {\doibase 10.1145/309739.309744} {\bibfield  {journal}
  {\bibinfo  {journal} {SIGACT News}\ }\textbf {\bibinfo {volume} {30}},\
  \bibinfo {pages} {33–35} (\bibinfo {year} {1999})}\BibitemShut {NoStop}%
\bibitem [{\citenamefont {Chau}\ and\ \citenamefont
  {Hardwick}(1998)}]{chau:1998}%
  \BibitemOpen
  \bibfield  {author} {\bibinfo {author} {\bibfnamefont {P.-L.}\ \bibnamefont
  {Chau}}\ and\ \bibinfo {author} {\bibfnamefont {A.}~\bibnamefont
  {Hardwick}},\ }\href {\doibase 10.1080/002689798169195} {\bibfield  {journal}
  {\bibinfo  {journal} {Mol. Phys.}\ }\textbf {\bibinfo {volume} {93}},\
  \bibinfo {pages} {511} (\bibinfo {year} {1998})}\BibitemShut {NoStop}%
\bibitem [{\citenamefont {Mehta}\ \emph {et~al.}(2019)\citenamefont {Mehta},
  \citenamefont {Bukov}, \citenamefont {Wang}, \citenamefont {Day},
  \citenamefont {Richardson}, \citenamefont {Fisher},\ and\ \citenamefont
  {Schwab}}]{mehta:2019}%
  \BibitemOpen
  \bibfield  {author} {\bibinfo {author} {\bibfnamefont {P.}~\bibnamefont
  {Mehta}}, \bibinfo {author} {\bibfnamefont {M.}~\bibnamefont {Bukov}},
  \bibinfo {author} {\bibfnamefont {C.-H.}\ \bibnamefont {Wang}}, \bibinfo
  {author} {\bibfnamefont {A.~G.}\ \bibnamefont {Day}}, \bibinfo {author}
  {\bibfnamefont {C.}~\bibnamefont {Richardson}}, \bibinfo {author}
  {\bibfnamefont {C.~K.}\ \bibnamefont {Fisher}}, \ and\ \bibinfo {author}
  {\bibfnamefont {D.~J.}\ \bibnamefont {Schwab}},\ }\href {\doibase
  https://doi.org/10.1016/j.physrep.2019.03.001} {\bibfield  {journal}
  {\bibinfo  {journal} {Physics Reports}\ }\textbf {\bibinfo {volume} {810}},\
  \bibinfo {pages} {1 } (\bibinfo {year} {2019})}\BibitemShut {NoStop}%
\bibitem [{\citenamefont {Vaart}(1998)}]{vaart:1998}%
  \BibitemOpen
  \bibfield  {author} {\bibinfo {author} {\bibfnamefont {A.~W. v.~d.}\
  \bibnamefont {Vaart}},\ }\href {\doibase 10.1017/CBO9780511802256} {\emph
  {\bibinfo {title} {Asymptotic Statistics}}},\ Cambridge Series in Statistical
  and Probabilistic Mathematics\ (\bibinfo  {publisher} {Cambridge University
  Press},\ \bibinfo {year} {1998})\BibitemShut {NoStop}%
\end{thebibliography}%
\let\addcontentsline\oldaddcontentsline% Restore \addcontentsline

\clearpage

\newcommand{\ra}[1]{\renewcommand{\arraystretch}{#1}}
\def\thetable{S\arabic{table}}
\def\thesection{S\arabic{section}}
\def\thesubsection{S\arabic{section}.\arabic{subsection}}
\def\thefigure{S\arabic{figure}}
\def\theequation{S\arabic{equation}}
\setcounter{figure}{0}
\setcounter{equation}{0}

\onecolumngrid

%SM
\begin{center}
	\textbf{\large Interpretable and unsupervised phase classification: Supplemental Material}\\[.5cm]
	Julian Arnold,$^{1}$ Frank Sch\"afer,$^{1}$ Martin \v{Z}onda,$^{2}$ and Axel U. J. Lode$^{2}$\\[.1cm]
	{\itshape ${}^1$Department of Physics, University of Basel, Klingelbergstrasse 82, 4056 Basel, Switzerland\\
		${}^2$Institute of Physics, Albert-Ludwigs-Universit\"at Freiburg,\\ Hermann-Herder-Strasse 3, 79104 Freiburg im Breisgau, Germany}\\
	(Dated: \today)\\[1cm]
\end{center}

This Supplemental Material contains technical details necessary for a reproduction of the
results presented in the main text and additional arguments to support our conclusions. We start by addressing details of the simulated annealing procedure to generate ground-state configurations of the Falicov-Kimball model (FKM). We provide technical details on the training and architecture of neural networks in the prediction-based method, as well as a derivation of its optimal predictive model in both the noise-free and noisy case. Next, order parameters and correlation functions quantifying the presence of different orderings and correlations in the FKM are defined. To contextualize the prediction-based method and motivate the mean-based method, we compare the methods to two alternative phase classification schemes. Next, technical details on the calculation of the indicator in the mean-based method are given and potential variations and extensions are discussed. Finally, we provide complementary figures (Fig.~\ref{fig_results_vector_field}-\ref{fig_fig3_different_inputs}) which support the discussion from the main text: the vector field arising in the prediction-based method, the inferred two-dimensional phase diagram using the mean-based method with $\abs{\bm{\mathcal{F}}_{0}}$ as input, as well as an analysis of the line-scan at $\rho=63/400\approx 0.16$ for different types of inputs in the noise-free and noisy case. The code for the prediction- and mean-based method that was utilized in this work is open source~\cite{github}.
\tableofcontents

\newpage
\section{Simulated annealing}\label{sec_SM_S1}
\indent For a fixed $f$-particle configuration $\bm{w}$, the Hamiltonian of the FKM in Eq.~\eqref{eq_fkm} in the main text can be transformed into   
\begin{equation}
{\cal H}^{\bm{w}} =  \sum_{j,j'} h_{jj'}d_{j}^{{\dagger}} d_{j'}^{\phantom{\dagger}} 
= \sum_\alpha \lambda^{\bm{w}}_\alpha b_\alpha^\dagger b_\alpha^{\phantom{\dagger}}, \label{eq:CentralHamiltonianSimpl}
\end{equation}
where we introduce the matrix elements $h_{jj'}=Uw_j\delta_{jj'}-t \delta_{\left\langle jj'\right\rangle}$. Its eigenvalues $\lambda_\alpha$ are obtained by numerical diagonalization. Finding the ground state of the FKM then means to find the configuration $\bm{w}$ which leads to the lowest energy 
\begin{equation}
E_{\mathrm{gs}}(\bm{w})=\sum_{\alpha=1}^{N_d}\lambda^{\bm w}_\alpha.  \label{eq:Egc}
\end{equation}

\indent However, even after accounting for the lattice symmetries, the ground-state configurations of systems with linear size $L=20$ can not be determined exactly by comparing the energies of all possible configurations $\bm{w}$ in general. An approximate method is required. Instead of using a reduced set of chosen orderings, as was done in previous studies~\cite{lemanski:2002,lemanski:2004} of the model, we determine the corresponding $f$-particle ground-state configuration $\bm{w}_{0}$ using simulated annealing.\\ 

\indent We use an algorithm based on a semi-classical Metropolis Monte Carlo~\cite{maska:2006}, where we use $E_{\mathrm{gs}}(\bm{w})$ in the statistical weights energy instead of the free energy. 
This means, that the candidate configuration $\bm{w}_c$, 
generated by a random displacement of a single $f$-particle from the current configuration $\bm{w}$, 
is accepted as new $\bm{w}$ if $E_{\mathrm{gs}}(\bm{w}_c) \leq E_{\mathrm{gs}}(\bm{w})$ or 
$\min(1,\exp[-\beta(E_{\mathrm{gs}}(\bm{w}_c)-E_{\mathrm{gs}}(\bm{w}))]) > r$, where $r$ is a random number drawn from a uniform distribution $r\in [0,1]$ and $\beta=1/T$ is the inverse temperature ($T$). We first used a classical protocol, where we started at relatively high temperature $T\sim0.1t$  and cooled the sample in $20-40$ discrete temperature steps to zero. A thermalization process consisting of $10^2-10^3 \times L^2$ updates was done at every time step.\\  

\indent However, we have found that an alternative adaptive protocol was much more efficient in lowering the energy. Namely, we start the annealing with a long thermalization at a low temperature (typically $T=0.003t$). In the next steps, depending on if the algorithm has found a configuration with lower energy at the current temperature or not, 
the temperature was either lowered by dividing its value by a factor between one and two (typically $1.25$) or increased by multiplying it by the same factor. The modified protocol is better in escaping local minima and has less troubles with the fact, that the FKM can go through more than one ordered phase with decreasing temperature~\cite{tran:2006,zonda:2012}.\\

\indent We have typically used a number of independent runs with random initial conditions. For small lattices ($L\leq10$) all simulations converged to configurations identical up to transformations of $p4m$. 
For $L=20$ we used $64$ runs with random initial conditions, plus several runs with initial configurations reflecting typical ground-state orderings identified for smaller lattices ($L\leq16$). We further investigated two cases in the main text. In the noisy case, we considered the $16$ configurations with the lowest energies. In the noise-free case, we performed one additional step: Namely, at each
investigated $\bm{p}=\left(U,\rho \right)$ we took the configuration $\bm{w}$ with the lowest energy and compared it with the energy calculated using the configuration $\bm{w}$ obtained as the ground state for the same $\rho$, but different (neighboring) $U$. The configuration $\bm{w}$ with the lowest energy was then taken as the final ground-state approximation.     

\section{Prediction-based method}\label{sec_SM_S2}
In this section, we provide details on the architecture and training of deep neural networks (DNNs) and linear models used in the prediction-based method. In particular, we list the corresponding hyperparameters employed throughout this work.

\subsection{Deep neural networks}
\indent In this work, we analyzed the ground-state phase diagram of the spinless FKM using the prediction-based method with DNNs as predictive models. These are built as follows: if the NN input is image-like, such as ground-state configurations $\bm{w}_{0}$ or the magnitude of their discrete Fourier transform $\abs{\bm{\mathcal{F}}_{0}}$, we first apply $K$ different square filters with the same linear size $L$ as the input image \cite{goodfellow:2016}. Subsequently, we apply a rectified linear unit, ${\rm ReLU}\left(x\right) = {\rm max}\left(0,x\right)$, as an activation function \cite{goodfellow:2016}. This results in an output feature map of size $1 \times 1 \times K$ which is then flattened to a feature vector with $K$ elements. In case of vector-like NN inputs, we skip this step. In both cases, we feed the corresponding vectors into a series of fully-connected layers (FCLs), where ReLUs are used as activation functions \cite{goodfellow:2016}. We note that the NN architecture remains to be optimized systematically. However, such a DNN with sufficiently many parameters can serve as an accurate predictive model.\\

\indent For training the DNNs, the inputs are standardized by means of the affine transformation
\begin{equation}\label{eq_standardization}
x_{i}' = \frac{x_{i} - \bar{x}_{i}}{\sigma_{i}},
\end{equation}
and the outputs are normalized as
\begin{equation}\label{eq_normalization}
x_{i}' = \frac{x_{i}}{\sigma_{i}},
\end{equation}
where $x_{i}$ denotes the $i$-th input/output, $\bar{x_{i}}$ and $\sigma_{i}$ are the mean and standard deviation of the distribution of the $i$-th input/output over the entire training data. Standardization ensures that the distribution of each transformed input $x_{i}'$ over the entire training data is characterized by $\left(\bar{x}'_{i}=0, \sigma'_{i}=1\right)$. Whereas normalization results in the distribution of each transformed output $x_{i}'$ over the entire training data being characterized by $\left(\bar{x}'_{i}=\bar{x}_{i}/\sigma_{i},\sigma'_{i}=1\right)$. Scaling of the inputs, here by means of standardization, is common practice in the data pre-processing step of machine learning tasks relying on gradient descent for optimization, as it generally yields a faster convergence rate \cite{lecun:2012}. The additional normalization of the outputs improves the model accuracy when training with a mean-square error (MSE) loss function, as it ensures that the outputs do not differ in size or spread and are consequently treated on an equal footing during the optimization. The MSE loss function is defined as
\begin{equation}\label{eq_mseloss}
\mathcal{L}_{\rm MSE} = \frac{1}{N_{\rm p}N_{\rm x}} \sum_{\bm{p}} \sum_{\bm{x}} \norm{\bm{p}-\hat{\bm{p}}\left(\bm{x}\right)}^2,
\end{equation}
where the sum runs over all $N_{\rm p}$ sampled points $\bm{p}$ in parameter space and all $N_{\rm x}$ inputs $\bm{x}$ at each point $\bm{p}$. Here, $\hat{\bm{p}}=\left(\hat{U},\hat{\rho} \right)$ denotes the predictions of the DNN given a particular input $\bm{x}$.\\

\indent The DNNs are implemented in PyTorch \cite{paszke:2019}, where the weights and biases are optimized using the stochastic gradient-based optimizer Adam~\cite{kingma:2014} to minimize the loss function $\mathcal{L}_{\rm MSE}$ [Eq.~\eqref{eq_mseloss}] over a series of epochs. After each training epoch, the vector-field divergence
\begin{equation}
\nabla_{\boldsymbol{p}} \cdot \boldsymbol{\delta p} = \left.\frac{\partial \delta U}{\partial U}\right|_{\bm{p}} + \left.\frac{\partial \delta \rho}{\partial \rho}\right|_{\bm{p}}
\end{equation}
is calculated based on the predictions $\hat{\bm{p}}$ for each sampled point $\bm{p}$ in parameter space. This is done using the symmetric difference quotient
\begin{align}\label{eq_derivatives}
\left.\frac{\partial \delta U}{\partial U}\right|_{\bm{p}} &\approx \frac{\delta U\left(U+\Delta U, N_{f}\right) - \delta U\left(U-\Delta U, N_{f}\right)}{2 \Delta U},\\
\left.\frac{\partial \delta \rho}{\partial \rho}\right|_{\bm{p}} &\approx \frac{\delta \rho\left(U, \rho+\Delta \rho\right) - \delta \rho\left(U, \rho - \Delta \rho\right)}{2 \Delta \rho},\notag
\end{align}
where $\delta U = \hat{U} - U$ and $\delta \rho = \hat{\rho} - \rho$. The divergence is averaged over all inputs per point $\bm{p}$. The learning rate is reduced by a fixed factor $f_{\rm r}$ if the loss $\mathcal{L}_{\rm MSE}$ does not drop below a certain relative threshold value within a given number of epochs, referred to as ``patience''. Gradients are calculated using backpropagation. During training, weights and biases are updated batch-wise, i.e., during each epoch the entire training data is randomly split into batches of equal size. For each batch, the predictions and the resulting loss $\mathcal{L}_{\rm MSE}$ are calculated and the NN parameters are then updated accordingly. \\

To incorporate configurations related through transformations of $p4m$ we use online data augmentation, i.e., each time a configuration is revisited during training a random transformation of $p4m$ is performed. Note that if we use $\abs{\bm{\mathcal{F}}_{0}}$ as inputs, we do not need to consider any lattice translations. In the case of symmetry-invariant inputs, such as correlation functions, we do not need to apply any transformations beforehand. Data augmentation is crucial, as all configurations related through transformations of $p4m$ have the same energy. Therefore, data augmentation aims at removing physically irrelevant differences between configuration samples and enforces the NN to pick up on patterns in the input, which are equally present in all the transformed versions. The DNN hyperparameters employed in this work are collected in Tab.~\ref{fig_hyperparameters_DNN}. The color scale in Fig.~\ref{fig_fig2}(b) and (c) in the main text was cut off at -1 and -2, respectively, for better visualization. There were very few distinct points in parameter space with a divergence signal $\nabla_{\boldsymbol{p}} \cdot \boldsymbol{\delta p}$ below this cut-off.
\begin{table*}[tbh!]
	\centering\ra{1.3}
	\begin{tabular}{@{}l|ccc|cccccccc@{}}\hline\hline
		Figure&\ref{fig_fig2}(b) & \ref{fig_fig2}(c) & \ref{fig_fig3}(a) & \ref{fig_fig3_different_inputs}(a) & \ref{fig_fig3_different_inputs}(g)\\\hline\hline
		$K$ & 2048 & 2048 & 512 & - & 512\\
		FCL 1 & (2048,1024) & (2048,1024) & (512,256) & (30,512) & (512,256)\\
		FCL 2 & (1024,512) & (1024,512) & (256,64) & (512,256) & (256,64)\\
		FCL 3 & (512,512) & (512,256) & (64,1) & (256,64) & (64,1)\\
		FCL 4 & (512,256) & (256,2) & - & (64,1) & -\\
		FCL 5 & (256,2) & - & - & - & -\\
		$N_{\rm tot}$ & 3838722 & 3576066 & 353153 & 163713 & 353153\\
		\hline
		learning rate & 0.001 & 0.0001 & 0.0001 & 0.0001 & 0.0001\\
		batch size & 700 & 7000 & 35 & 35 & 35\\
		$f_{\rm r}$ & 0.5 & 0.5 & 0.5 & 0.5 & 0.5\\
		patience & 50 & 50 & 50 & 50 & 50\\
		epochs & 1576 & 1485 & 770 & 1057 & 893\\
		\hline\hline
	\end{tabular}
	\caption{DNN hyperparameters employed in this work. Here, the number of inputs $n_{\rm in}$ and outputs $n_{\rm out}$ of each fully-connected layer (FCL) is denoted as ($n_{\rm in}$, $n_{\rm out}$). The total number of NN parameters (weights and biases) is denoted as $N_{\rm tot}$. Default settings are used except where explicitly stated.}
	\label{fig_hyperparameters_DNN}
\end{table*}

\subsection{Linear models}\label{sec_linear_models}
To construct directly interpretable local order parameters for each predicted phase transition, we rely on the prediction-based method with linear models. The predictions $\hat{\bm{p}}$ of a linear model are given as
\begin{equation}
\hat{\bm{p}} = \theta \bm{x} + \bm{b},
\end{equation}
with an input vector $\bm{x}$, a weight matrix $\theta$ and an vector of additive biases $\bm{b}$. Evidently, a linear model allows for a direct interpretation in terms of its weights $\theta$ and biases $\bm{b}$. In addition to the MSE term in Eq.~\eqref{eq_mseloss}, when training linear models we add a $L_{2}$ regularization term to our loss function
\begin{equation}\label{eq_generalloss}
\mathcal{L} = \mathcal{L}_{\rm MSE} + \lambda \sum_{i} \theta_{i}^2.
\end{equation}
Here, the regularization rate $\lambda$ controls the strength of our preference for smaller weights $\theta_{i}$, where the sum runs over all weights \cite{goodfellow:2016}. In this work, $\lambda$ is chosen small enough such that the trained model qualitatively yields the same predictions as a model trained to minimize $\mathcal{L}_{\rm MSE}$. The additional restrictions posed on the model by the $L_{2}$ regularization term thereby removes the remaining freedom in its parameters. This is of particular importance when trying to interpret the model through its weights and biases \cite{blucher:2020, casert:2019, zhang:2020}.\\

\indent The linear models are trained to minimize $\mathcal{L}$ [Eq.~\eqref{eq_generalloss}] using the scikit-learn implementation for Ridge regression with default settings \cite{scikit:2011}. Data augmentation is performed offline by applying $n_{\rm trafo}$ random transformations of $p4m$ to each input configuration beforehand analogous to the online variant described previously. The hyperparameters for training the linear models employed in this work are collected in Tab.~\ref{fig_hyperparameters_linear}.
\begin{table*}[tbh!]
	\centering\ra{1.3}
	\begin{tabular}{@{}l|ccccc|cc|cccc@{}}\hline\hline
		Figure&\ref{fig_fig3}(b) & \ref{fig_fig3_different_inputs}(b) & \ref{fig_fig3_different_inputs}(h)\\\hline\hline
		$n_{\rm trafo}$ & 20 & 20 & 0\\
		$\lambda$ & 1 & 1 & $10^{-7}$\\
		\hline\hline
	\end{tabular}
	\caption{Hyperparameters for training the linear models employed in this work. Default settings were used except where explicitly stated.}
	\label{fig_hyperparameters_linear}
\end{table*}
\newpage
\subsection{Derivation of optimal predictive model}
In this section, we provide an analytical derivation of the optimal model predictions when using the prediction-based method for phase classification in the general noisy case, as well as the special noise-free case.\\ 
\begin{figure}[htb!]
	\begin{center}
		\includegraphics[width=0.7\textwidth]{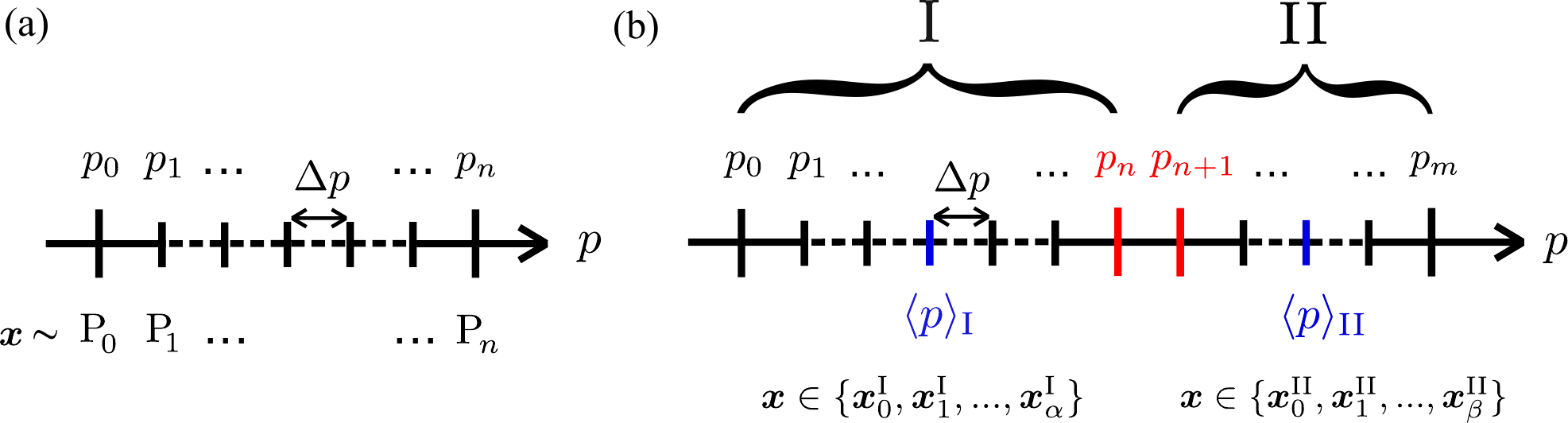}
		\caption{Schematic representation of a 1D space spanned by the parameter $p$ which is sampled equidistantly with a spacing $\Delta p$. (a) At each point $p_{i}$ samples $\bm{x}$ are drawn from a given underlying probability distribution $\bm{x} \sim {\rm P}_{i}\left( \bm{x} \right)$. (b) Two distinct regions, labelled I and II, in 1D space spanned by the parameter $p$. For all points $p \in \{p_{0},p_{1},...,p_{n} \}$ in region I samples are drawn from the same set $\bm{x} \in \{\bm{x}_{0}^{\rm I},\bm{x}_{1}^{\rm I},...,\bm{x}_{\alpha}^{\rm I} \}$. Similarly, for all points $p \in \{ p_{n+1},p_{n+2},...,p_{m} \}$ in region II samples are drawn from $\bm{x} \in \{\bm{x}_{0}^{\rm II},\bm{x}_{1}^{\rm II},...,\bm{x}_{\beta}^{\rm II} \}$. The points $p_{n}$ and $p_{n+1}$ (red) make up the boundary of the two regions and the center of mass of region I/II is denoted as $\langle p \rangle_{\rm I/II}$ (blue), respectively.}
		\label{fig_results_noisy_derivation}
	\end{center}
\end{figure}
\indent {\it Noisy case} -- We assume to have a system with a single, tunable parameter $p$, which we sample on an equidistant grid with a grid spacing $\Delta p$. At each grid point $p_{i}$, we draw inputs $\bm{x}$ from an underlying probability distribution $\bm{x} \sim {\rm P}_{i}\left( \bm{x} \right)$. This situation is illustrated in Fig.~\ref{fig_results_noisy_derivation}(a). We train a model $f: \bm{x} \rightarrow f\left(\bm{x}\right)$ to infer $p$ from the samples $\{ \bm{x} \}$ generated at $p$, i.e., to minimize a MSE loss function 
\begin{equation}
\mathcal{L}_{\rm MSE} = \frac{1}{N_{\rm p}N_{\rm x}} \sum_{p} \sum_{\bm{x}} \left(p - f\left(\bm{x}\right)\right)^2.
\end{equation}
Here, $p$ runs over all $N_{\rm p}$ sampled grid points and $\bm{x}$ runs over all $N_{\rm x}$ inputs at $p$.\\

Let us pick a particular input $\bm{x}_{j}$. We can determine the optimal model prediction $f_{\rm opt}\left(\bm{x}_{j}\right)$ by minimizing $\mathcal{L}_{\rm MSE}$ w.r.t. $f\left(\bm{x}_{j}\right)$, i.e.,
\begin{equation}
\frac{\partial \mathcal{L}_{\rm MSE}}{\partial f\left(\bm{x}_{j}\right)} = \frac{2}{N_{\rm p}N_{\rm x}} \sum_{p} N_{{\rm x}}^{j} \left( p \right) \left(p - f_{\rm opt}\left(\bm{x}_{j}\right)\right) = 0.
\end{equation}
Here, $N_{\rm x}^{j}\left( p_{i} \right)$ denotes the number of times the particular input $\bm{x}_{j}$ is present at point $p_{i}$, and ${\rm P}_{i}\left( \bm{x}_{j}\right) \equiv N_{\rm x}^{j}\left( p_{i} \right)/N_{\rm x}$ is the associated probability. We can additionally define ${\rm \tilde{P}}_{i}\left( \bm{x}_{j}\right) \equiv {\rm P}_{i}\left( \bm{x}_{j}\right)/ \sum_{p} {\rm P}\left(\bm{x}_{j}\right)$ as the probability of drawing the particular input $\bm{x}_{j}$ at point $p_{i}$ compared to all other sampled points $p$. We then obtain
\begin{equation}\label{eq_noisy_optimalpredictions}
f_{\rm opt}\left(\bm{x}_{j}\right) = \sum_{p} {\rm \tilde{P}}_{i}\left( \bm{x}_{j}\right) p.
\end{equation}
Repeating this step for all inputs $\{ \bm{x} \}$, we find that any model $f_{\rm opt}$ which minimizes $\mathcal{L}_{\rm MSE}$ will output $f_{\rm opt}\left(\bm{x}_{j}\right)$ for an input $\bm{x}_{j}$. This implies that the optimal model predicts the center of mass for a particular input $\bm{x}_{j}$, where each grid point is weighted according to the probability to draw the input $\bm{x}_{j}$. Note that there are no additional restriction on the form of $f$.\\ 

Given an optimal model $f_{\rm opt}$, the divergence of $\delta p$ at a point $p_{i}$ is calculated as
\begin{equation}
\left.\frac{\partial \delta p}{\partial p}\right|_{p_{i}} \approx \frac{\delta p\left(p_{i} + \Delta p\right) - \delta p\left(p_{i} - \Delta p\right)}{2 \Delta p}.
\end{equation}
Here, $\delta p = \hat{p}- p$ with $\hat{p} = \frac{1}{N_{\rm x}} \sum_{\bm{x}} f_{\rm opt}\left(\bm{x}\right)$, where the sum runs over all $N_{\rm x}$ samples $\{ \bm{x} \}$ drawn at point $p$. Hence,
\begin{equation}\label{eq_divergence_derived}
\left.\frac{\partial \delta p}{\partial p}\right|_{p_{i}} \approx \frac{\hat{p}\left(p_{i} + \Delta p\right) - \hat{p}\left(p_{i} - \Delta p\right)}{2 \Delta p}-1.
\end{equation}
The generalization to a parameter space of arbitrary dimension is straightforward.\\

\indent Here, we have derived the divergence signal of an optimal model $f_{\rm opt}$ for the most general (noisy) situation. This removes the need for further interpretation of the DNN because it merely serves to approximate $f_{\rm opt}$, and thereby renders the method interpretable. Additionally, it opens up the possibility to approximate $f_{\rm opt}$ without the need of DNNs as universal function approximators. Specifically, one may compute the optimal model predictions $f_{\rm opt}$ in Eq.~\eqref{eq_noisy_optimalpredictions} from estimates of the underlying probability distributions ${\rm P}\left(\bm{x} \right)$, e.g., obtained using Monte Carlo methods. A comparison of such approaches to NN-based ones will be subject to future studies.\\ 

\indent Note that even with the form of the optimal model $f_{\rm opt}$ at hand, we still may want to investigate the decision-making of DNNs trying to approximate $f_{\rm opt}$ to obtain physical insights, e.g., using state-of-the-art attribution methods following Ref.~\cite{wetzel:2017, casert:2019}. However, the scheme based on training linear models for predicted phase transitions to construct local order parameters proposed in the main text proved to be more effective to automate this task.\\

\indent {\it Noise-free case} -- Consider now the special noise-free case in which there are regions along $p$ where samples $\{ \bm{x} \}$ are identical up to transformations of $p4m$. A situation with two such regions, labelled I and II is shown in Fig.~\ref{fig_results_noisy_derivation}(b). In particular, there is a single set $\bm{x} \in \{\bm{x}_{0}^{\rm{I}},\bm{x}_{1}^{\rm{I}},...,\bm{x}_{\alpha}^{\rm{I}} \}$ from which the samples are drawn at each point $p \in \{ p_{0},p_{1},...,p_{n} \}$ within region I. Similarly, samples at all points $p \in \{ p_{n+1},p_{n+2},...,p_{m} \}$ within region II are drawn from set $\bm{x} \in \{\bm{x}_{0}^{\rm{II}},\bm{x}_{1}^{\rm{II}},...,\bm{x}_{\beta}^{\rm{II}} \}$ not related by transformations of $p4m$ to the set of region I. Consequently, ${\rm \tilde{P}}\left(\bm{x}_{j}^{I}\right)=0$ for all points $p$ in region II and ${\rm \tilde{P}}\left(\bm{x}_{j}^{I}\right)={\rm const.}=1/N_{\rm p}^{\rm I}$ for all points $p$ in region I, and vice-versa. Here, $N_{\rm p}^{\rm I}$ denotes the number of sampled points in region I. Using Eq.~\eqref{eq_noisy_optimalpredictions} we then obtain
\begin{equation}
f_{\rm opt}\left(\bm{x}_{i}^{\rm I}\right) = \frac{1}{N_{\rm p}^{\rm I}} \sum_{p \in \: {\rm I}} p =\langle p \rangle_{\rm I},
\end{equation}
where $p$ runs over all $N_{\rm p}^{\rm I}$ sampled points in region I. Similarly, $f_{\rm opt}\left(\bm{x}_{i}^{\rm II}\right) =\langle p \rangle_{\rm II}$. Meaning, that the optimal model yields predictions at the center of mass for all samples within a given region.\\

\indent Given an optimal model $f_{\rm opt}$, the divergence of $\delta p$ at a point $p_{i}$ is given by Eq.~\eqref{eq_divergence_derived}. Hence, for any point $p_{i}$ \emph{within region} $j \in \{ {\rm I, II} \}$ that is not directly located in vicinity of the boundary of the region, we have
\begin{equation}\label{eq_withinphase}
\left.\frac{\partial \delta p}{\partial p}\right|_{p_{i}} \approx -1.
\end{equation}
Conversely, for the two points \emph{at the boundary} between region I and II we have
\begin{equation}\label{eq_phaseboundary}
\left.\frac{\partial \delta p}{\partial p}\right|_{p_{i}} \approx \frac{\langle p \rangle_{\rm II} - \langle p \rangle_{\rm I}}{2\Delta p}-1.
\end{equation}
This shows that the prediction-based method predicts a phase transition whenever neighboring inputs cannot be related through transformations of $p4m$. Equivalently, it predicts phases to be regions in which neighboring inputs are simply related by transformations of $p4m$. The value of the divergence peak at a phase boundary serves as an indicator of the mean extent of the corresponding phases in parameter space. The size of a phase is thus linked to its stability, i.e., its robustness against variations in the system parameters. This procedure can be generalized straightforwardly to a parameter space of arbitrary dimension with an arbitrary number of phases.\\

\begin{figure}[htb!]
	\begin{center}
		\includegraphics[width=0.5\textwidth]{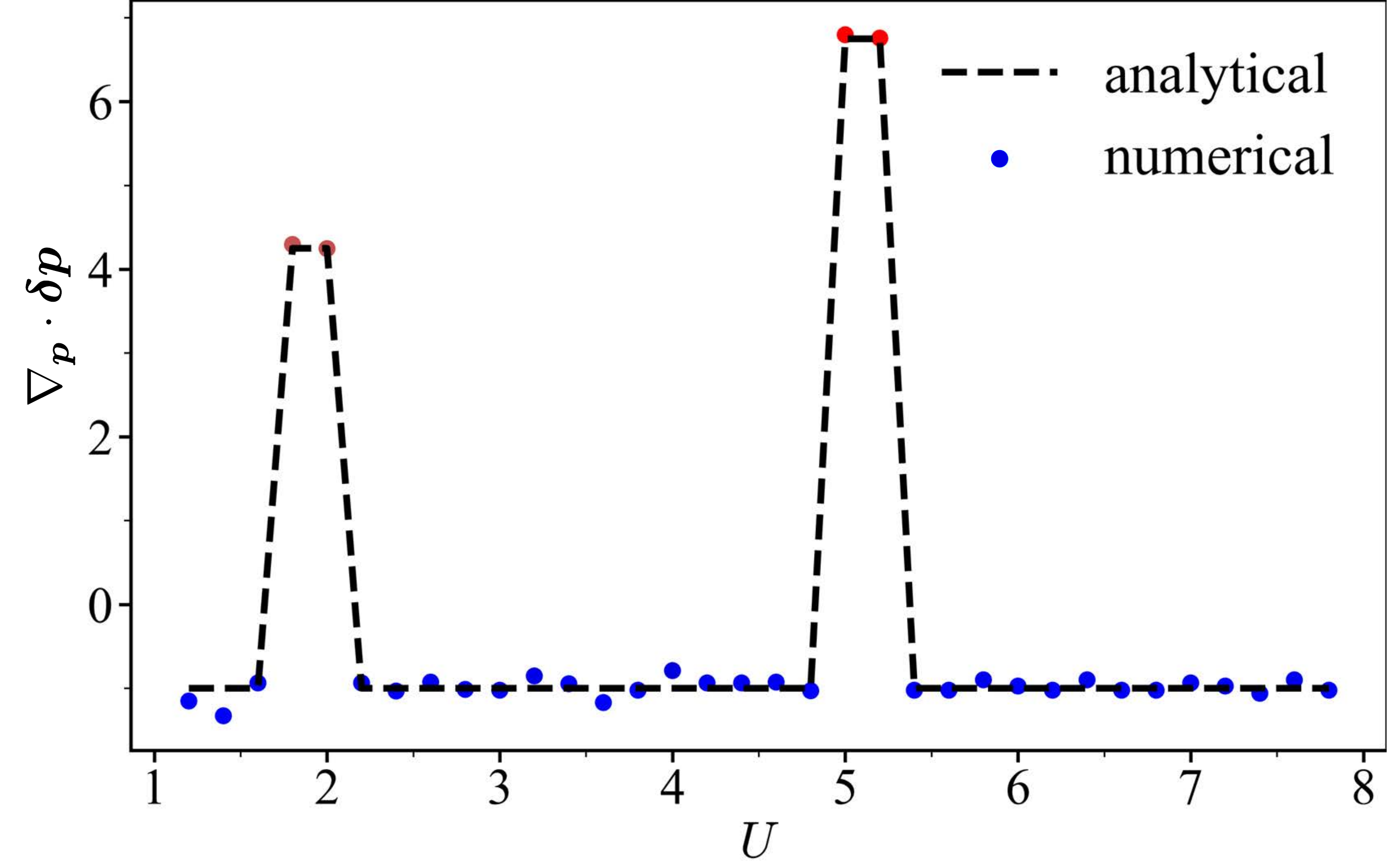}
		\caption{Vector-field divergence $\nabla_{\bm{p}} \cdot \bm{\delta p}$ as a function of $U$ at $\rho=35/400$ obtained analytically based on Eq.~\eqref{eq_withinphase} and \eqref{eq_phaseboundary}, as well as numerically using a DNN trained with $\abs{\bm{\mathcal{F}}_{0}}$ as input for the two-dimensional ground-state phase diagram in the ''noise-free`` case [see Fig.~\ref{fig_fig2}(b) in the main text for full predicted phase diagram].}
		\label{fig_comparison_analytica_numerical}
	\end{center}
\end{figure}

\indent Figure~\ref{fig_comparison_analytica_numerical} shows the vector-field divergence $\nabla_{\bm{p}} \cdot \bm{\delta p}$ as a function of $U$ at $\rho=35/400$ for the DNN trained using $\abs{\bm{\mathcal{F}}_{0}}$ as input in the noise-free case whose predicted two-dimensional ground-state phase diagram of the FKM is shown in Fig.~\ref{fig_fig2}(b) in the main text. The values of the divergence match the results in Eq.~\eqref{eq_withinphase} and \eqref{eq_phaseboundary} almost perfectly. This confirms that our trained predictive model is indeed optimal, i.e., minimizes $\mathcal{L}_{\rm MSE}$.\\

\section{Order parameters and correlation functions}\label{sec_SM_S3}
In this section, we discuss order parameters for segregated, diagonal and axial orderings [cf. labels (\underline{1})-(\underline{3}) in Fig.~\ref{fig_fig2} in the main text], and provide definitions for set of three correlation functions $\kappa^{\xi}_{n}$ measuring square, axial, and diagonal order.

\subsection{Order parameters}
\begin{figure}[htb!]
	\begin{center}
		\includegraphics[width=0.2\textwidth]{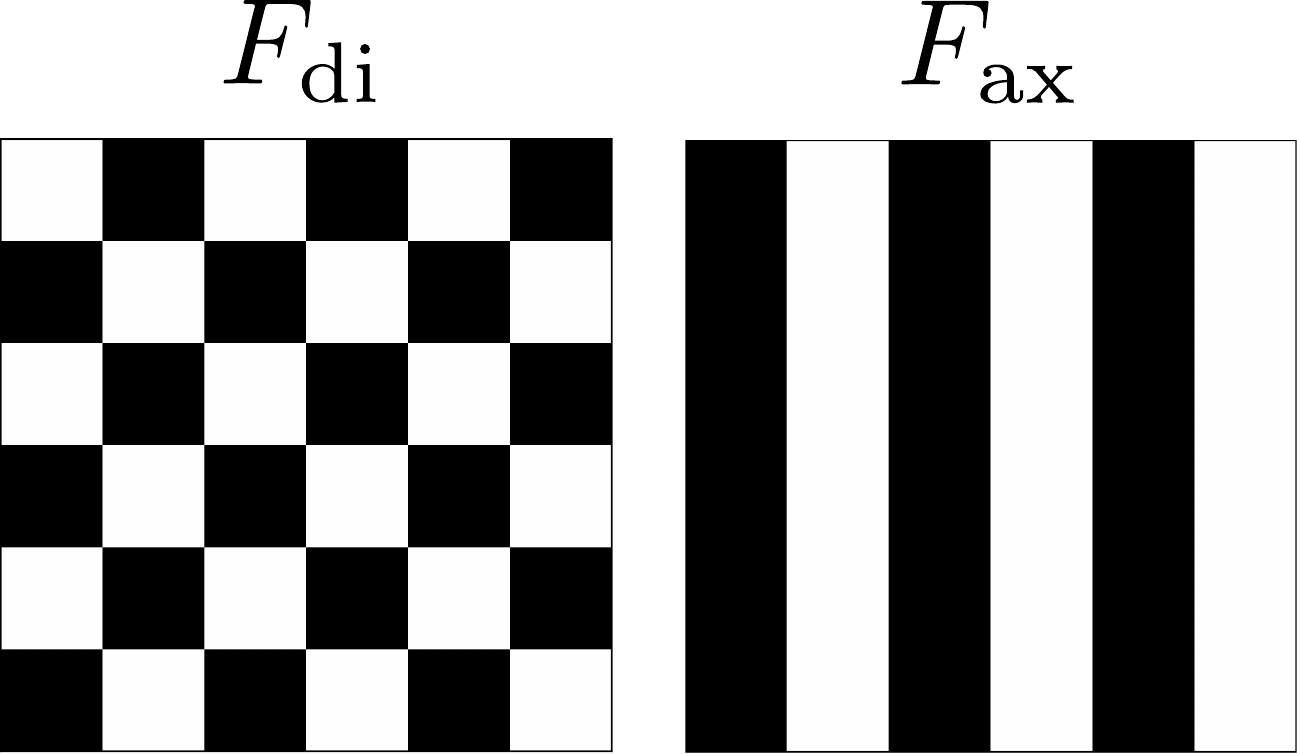}
		\caption{Representative filter $F_{\xi}$ for computing the order parameter for diagonal ($F_{\rm di}$) or axial ($F_{\rm ax}$) orderings. Here, black denotes 1 and white denotes -1.}
		\label{fig_results_op_filters}
	\end{center}
\end{figure}

\indent To define order parameters for the diagonal (di), as well as axial (ax) orderings, we introduce appropriate filters $F_{\xi},\: \xi \in \{ {\rm di, ax} \}$. The values of the order parameters are obtained by taking the Frobenius scalar product of the raw configurations $\bm{w}$ with the corresponding filters. To account for configurations that are related through transformations of $p4m$, we also subject the filters to the corresponding transformations. Ultimately, we take the maximum value over all symmetry-related filters $\{ F_{\xi} \}$ as the value of the order parameter $O_{\xi}$ for a particular configuration sample $\bm{w}$:
\begin{equation}
O_{\xi}\left( \bm{w} \right)\equiv\max_{F_{\xi}} \frac{1}{L^2} \sum_{i=1}^{L} \sum_{j=1}^{L} \left(F_{\xi} \odot \left( 2\bm{w}-\mathbb{1}\right)\right)_{ij},
\end{equation}
where $\odot$ denotes the element-wise product and $\mathbb{1}$ is the identity matrix. Figure~\ref{fig_results_op_filters} displays representative filters for the order parameters of diagonal and axial orderings, where all other filters can be obtained from these examples through transformations of $p4m$. Note that the filters have the same size as the configurations they are applied to, here $L=20$. The filters for lattices of different size can be defined analogously by retaining the same patterns as in Fig.~\ref{fig_results_op_filters}. If necessary, order parameters for other orderings [cf. labels (\underline{4})-(\underline{9}) in Fig.~\ref{fig_fig2} in the main text] can be defined in a similar manner.\\

\indent Defining an order parameter for the segregated (sg) ordering is conceptually simple. It amounts to determining whether the configuration sample contains a single, connected cluster of $f$ particles. This is implemented by a backtracking algorithm \cite{kreher:1999}. We define a binary order parameter $O_{\rm sg}$ taking on the value 1 (0) if the configuration sample does (not) contains a single, connected cluster of $f$ particles (as determined by the algorithm).\\

\indent Clearly, all three order parameters $O_{\xi},\: \xi \in \{ {\rm di, ax, sg} \}$ share a common set of desired properties \cite{chau:1998}. In particular, the order parameters are invariant under when the input configuration samples are subjected to transformations of $p4m$. Furthermore, the maximum value of the order parameters is $\max_{ \bm{w}} O_{\xi}\left( \bm{w} \right) = 1$ which is only achieved for samples showing perfect ordering of type $\xi$. Additionally, the three order parameters defined by means of filters can take on values ranging from 0 to 1, indicating the partial presence of the corresponding pattern.\\

\begin{figure}[htb!]
	\begin{center}
		\includegraphics[width=0.99\textwidth]{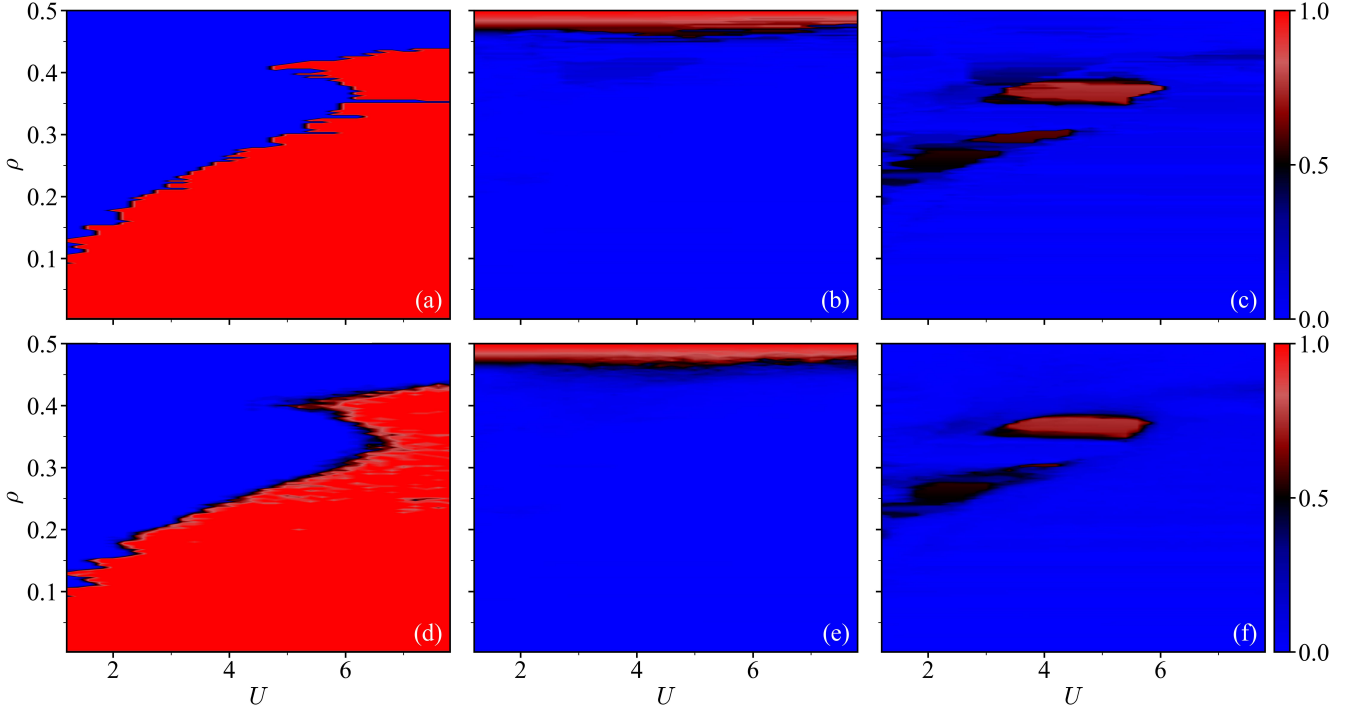}
		\caption{Values of order parameters (a),(d) $O_{\rm sg}$, (b),(e) $O_{\rm ax}$, and (c),(f) $O_{\rm di}$ in the (a)-(c) noise-free and (d)-(f) noisy case.}
		\label{fig_results4}
	\end{center}
\end{figure}

\indent Figure~\ref{fig_results4} shows the values of all three order parameters $O_{\xi},\: \xi \in \{ {\rm di, ax, sg} \}$ for each sampled point $\bm{p} = \left(U, \rho \right)$ in parameter space for the FKM in the noise-free and noisy case. In the noisy case, we additionally average over all available configurations at each point $\bm{p}$ to obtain a scalar value. The order parameters reveal the presence of segregated, diagonal, and axial orderings marked as (\underline{1}), (\underline{2}), and (\underline{3}) in Fig.~\ref{fig_fig2} in the main text, respectively.
\newpage
\subsection{Correlation functions}
\indent As a set of observables derived from the configurations which remain invariant under transformations of $p4m$ we here consider a minimal set of three different correlation functions $\kappa_{n}^{\xi}, \: \xi \in \{ {\rm sq, ax, di} \}$. In particular, these measure square ($\kappa_{n}^{\rm sq}$), axial ($\kappa_{n}^{\rm ax}$), and diagonal correlations ($\kappa_{n}^{\rm di}$) over $n$ lattice sites. For a configuration sample $\bm{w}$, the three correlation functions are calculated as
\begin{align}\label{eq_corr}
\kappa_{n}^{\rm sq} &= \frac{1}{8nL^2} \sum_{i=1}^{L} \sum_{j=1}^{L}\sum_{\alpha=-n}^{n} \sum_{\beta=-n}^{n} \left(\delta_{|\alpha|,n} + \delta_{|\beta|,n} - \delta_{|\alpha|+|\beta|,2n} \right) \left(2w_{i,j}-1\right)\left(2w_{i+\alpha,j+\beta}-1\right),\\
\kappa_{n}^{\rm ax} &= \frac{1}{4L^2} \sum_{i=1}^{L} \sum_{j=1}^{L}\sum_{\alpha \in \{-n,n\}} \sum_{\beta \in \{-n,n\}} \delta_{|\alpha|+|\beta|,n} \left(2w_{i,j}-1\right)\left(2w_{i+\alpha,j+\beta}-1\right),\notag\\
\kappa_{n}^{\rm di} &= \frac{1}{4L^2} \sum_{i=1}^{L} \sum_{j=1}^{L}\sum_{\alpha \in \{-n,n\}} \sum_{\beta \in \{-n,n\}} \delta_{|\alpha|+|\beta|,2n} \left(2w_{i,j}-1\right)\left(2w_{i+\alpha,j+\beta}-1\right).\notag
\end{align}
Since we assume periodic boundary conditions, the largest unique $n$ is given by $n = L/2$. The computation of the correlation functions from Eq.~\eqref{eq_corr} is illustrated in Fig.~\ref{fig_fig2}(d) of the main text. In case of the FKM on a square lattice, we find that these three correlation functions and combinations thereof are sufficient to describe most patterns and ordering. Therefore, they represent a physically-motivated way of detecting phase transitions based on the magnitude of the change in order in the set of correlation functions.

\section{Alternative phase classification methods}\label{sec_SM_S4}
The derivation of the optimal model predictions has strengthened our understanding of the prediction-based method. While we have so far relied on DNNs to approximate $f_{\rm opt}$, it opens up the possibility of devising algorithms which yield the same predictions as $f_{\rm opt}$, or equivalently, reproduce the predicted phase diagram. To reproduce the predicted ground-state phase diagram in the noise-free case obtained using the prediction-based method [see Fig.~\ref{fig_fig2}(b) in the main text], a model simply needs to be sensitive to any change in the configurations (up to transformations of $p4m$). If a change is detected, a new phase is declared. Once all the points in parameter space are analysed, the full phase diagram is obtained. The main approach we take in this work is to adopt a representation of the configurations which is invariant under transformations of $p4m$: the correlation functions in Eq.~\eqref{eq_corr}. Consequently, a simple comparison of the representations reveals the phase diagram. Furthermore, the extension of this approach to the noisy case is straightforward and eventually leads to the formulation of the general mean-based method. In what follows, we describe two alternative approaches to reproduce the results of the prediction-based method in the special noise-free case.\\

\indent As a naïve first approach, one could compare the ground-state configuration samples of different points in parameter space. If a symmetry transformation is found which relates the configuration samples, the corresponding points belong to the same phase, otherwise a new phase is declared. Clearly, the computational complexity of such an approach is reduced significantly by considering $\abs{\bm{\mathcal{F}}_{0}}$, as opposed to $\bm{w}_{0}$ to remove the need to consider lattice translations. Note that such an approach fails when considering the general noisy case.\\

\indent As a second approach, we propose to use the system Hamiltonian as motivated by the simulated annealing procedure. For a given point in parameter space (point I), we take the corresponding ground-state configuration sample and calculate its energy using the system Hamiltonian at a neighbouring point along $U$ (point II). Additionally, we evaluate the energy of the ground-state configuration sample at point II using the system Hamiltonian at point II. If the difference in energy is smaller than an appropriate threshold value, we may regard the two samples degenerate and assign them to the same phase. This is valid, since both samples are equally likely to be generated using the simulated annealing procedure. Otherwise a new phase is declared. Analysing the entire two-dimensional parameter space in this fashion will eventually yield the same phases as predicted by the prediction-based method. Note that such an approach assumes that we have perfect knowledge of the system Hamiltonian, which is usually not the case in experiments. Furthermore, an extension of this approach to the general noisy case may not be straightforward.

\section{Mean-based method}\label{sec_SM_S5}
As a key result, we propose the mean-based method as a novel data-driven scheme for identifying and characterizing phase transitions in an automated fashion. It relies on a difference signal $\Delta \bar{x}$ that serves as a generic indicator for phase transitions which is calculated as
\begin{equation}\label{eq_differencemethod}
\Delta \bar{x} \left( \bm{p} \right) = \norm{\bar{\bm{x}}\left(U+\Delta U, \rho\right) - \bar{\bm{x}}\left(U-\Delta U, \rho\right)}.
\end{equation}
Here, $\bar{\bm{x}}\left(\bm{p}_{i} \right)= \sum_{j} P_{i}\left( \bm{x}_{j} \right) \bm{x}_{j} \approx 1/N_{\rm x}^{j} \sum_{j} \bm{x}_{j}\left(\bm{p}_{i} \right)$ denotes the average input at a point $\bm{p}_{i}$, where we average over all corresponding $N_{\rm x}^{j}$ input $\bm{x}_{j}$.\\

\indent Given a set of configurations $\{\bm{x} \}$ at a point $\bm{p}_{i}$ we perform offline data augmentation (see training procedure for DNNs and linear models in Section~\ref{sec_SM_S2}) by applying $n_{\rm trafo}$ random transformations of $p4m$ to each configuration. Based on the augmented set of configurations, we then compute input features $\bm{x}$ (such as $\bm{\kappa}$ or $\abs{\bm{\mathcal{F}}_{0}}$) or use the configurations themselves to obtain the corresponding average $\bar{\bm{x}}$. In this work, we choose $n_{\rm trafo}=20,0$ when considering the magnitude of their discrete Fourier transform $\abs{\bm{\mathcal{F}}}$, and corresponding correlation functions $\bm{\kappa}$, respectively. The difference in $n_{\rm trafo}$ when using different inputs is based on the fact, that a reduced number of transformations need to be considered when the chosen input is invariant under (parts of) the transformations of $p4m$, as is the case for $\abs{\bm{\mathcal{F}}}$ or $\bm{\kappa}$. Clearly, this results in reduced computational cost compared to using an input which is (in general) not invariant under transformations of $p4m$, such as the raw configurations $\bm{w}$ themselves.\\

\indent Note that the generic indicator in Eq.~\eqref{eq_differencemethod} can easily be extended to include changes in $\rho$. This can, for example, be accomplished by defining the generic indicator as
\begin{equation}\label{eq_differencemethod_3}
\Delta \bar{x} \left( \bm{p} \right) = \norm{\bar{\bm{x}}\left(U+\Delta U, \rho\right) - \bar{\bm{x}}\left(U-\Delta U, \rho\right)} + \norm{\bar{\bm{x}}\left(U, \rho+\Delta \rho \right) - \bar{\bm{x}}\left(U, \rho-\Delta \rho \right)}.
\end{equation}
Similarly, we may extend the indicator to parameter spaces of arbitrary dimension.\\

\indent Clearly, the mean-based method relying on Eq.~\eqref{eq_differencemethod} as an indicator fails at identifying phase transitions for which $\bar{\bm{x}}$ remains unchanged. As different inputs will be more suitable depending on the nature of the phase transitions under investigation, an appropriate choice of input can likely combat such failures. However, we aim to provide a method which does not require significant tuning of the input, e.g., based on incorporating physical knowledge of the system at hand. For this, one may extend the approach to detect changes in the $m$-th order moments $\mu_{x,m}$ of the underlying probability distributions ${\rm P}\left(\bm{x} \right)$ as opposed to the mean. The corresponding indicators are then given as
\begin{equation}\label{eq_differencemethod_2}
\Delta \mu_{x,m} \left( \bm{p} \right) = \norm{\bm{\mu}_{x,m}\left(U+\Delta U, \rho\right) - \bm{\mu}_{x,m}\left(U-\Delta U, \rho\right)}.
\end{equation}
Equation~\eqref{eq_differencemethod_2} represents variants of the mean-based method relying on different measures that characterize changes in the underlying probability distributions ${\rm P}\left(\bm{x} \right)$. In particular, such indicators may yield complementary information about the corresponding phase transitions.\\ 

\indent We continue the thought of devising measures which quantify changes in the distributions ${\rm P}\left(\bm{x} \right)$, and can thereby serve as indicators for phase transitions, by considering the Hellinger distance $H\left(P_{i},P_{j} \right)$ \cite{vaart:1998} as a measure for the similarity between two probability distributions $P_{i}$ and $P_{j}$. The corresponding indicator $I\left(\bm{p} \right)$ for phase transitions would then simply be given as $I\left(\bm{p} \right)\equiv H\left(P_{\left(U+\Delta U,\rho \right)},P_{\left(U-\Delta U,\rho \right)} \right)$. In future studies, such extensions of the mean-based method should be investigated and compared based on the insights into the phase diagrams they provide and their computational cost.
\section{Complementary figures}\label{sec_SM_S6}
\begin{figure*}[htb!]
	\begin{center}
		\includegraphics[width=0.9\textwidth]{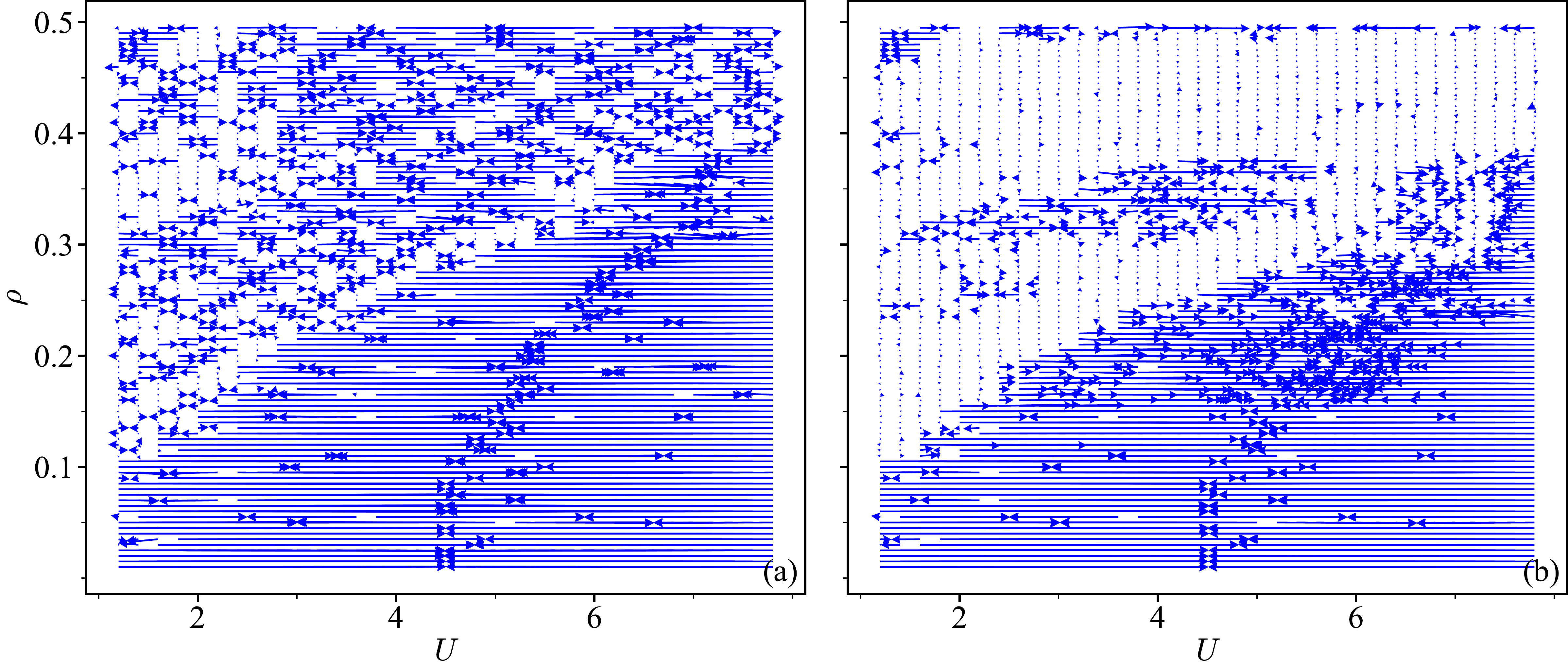}
		\caption{Vector field $\bm{\delta p}=\hat{\bm{p}}-\bm{p}$ obtained using a DNN trained with $\abs{\bm{\mathcal{F}}_{0}}$ as input for the two-dimensional phase diagram of the FKM in (a) the noise-free and (b) noisy case [see Fig.~\ref{fig_fig2}(b),(c) in the main text for predicted phase diagrams, respectively]. The vector field exhibits a horizontal structure which demonstrates that $\rho$ is predicted with near-perfect accuracy.}
		\label{fig_results_vector_field}
	\end{center}
\end{figure*}

\begin{figure*}[htb!]
	\begin{center}
		\includegraphics[width=0.9\textwidth]{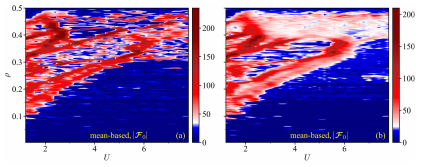}
		\caption{Indicator $\Delta \bar{x}$ [Eq.~\eqref{eq_differencemethod}] based on $\abs{\bm{\mathcal{F}}_{0}}$ as input on the two-dimensional parameter space of the FKM in (a) the noise-free and (b) the noisy case. This shows that the mean-based approach also works when using $\abs{\bm{\mathcal{F}}_{0}}$, as opposed to $\bm{\kappa}$ as input [cf. Fig.~\ref{fig_fig2}(e),(f) in the main text].}
		\label{fig_results_inputdifference_noisefree_noisy}
	\end{center}
\end{figure*}

\begin{figure*}[htb!]
	\begin{center}
		\includegraphics[width=0.99\textwidth]{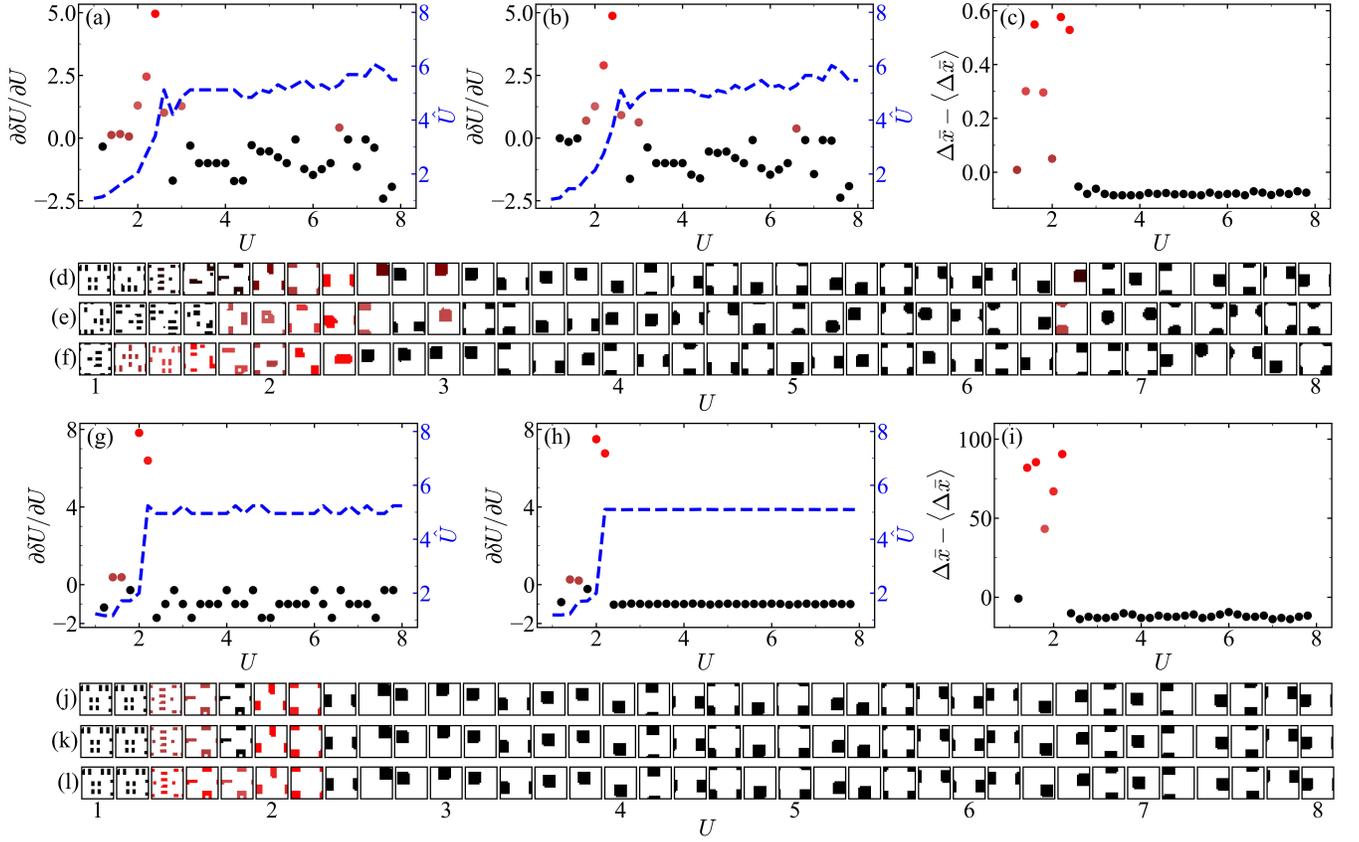}
		\caption{Analysis of the transition from non-segregated to segregated orderings occurring along the line-scan from $U_{\rm min}=1$ to $U_{\rm max}=8$ at fixed $\rho=63/400\approx0.16$ [cf. dashed line in Fig.~\ref{fig_fig2}(c),(f) in the main text] in (a)--(f) the noisy and (g)--(l) noise-free case. (a)-(f) Predictions $\hat{U}$ and corresponding divergence $\partial \delta U/\partial U$ of (a),(d) a DNN and (b),(e) a linear model, as well as (c),(f) the indicator $\Delta \bar{x}$ [Eq.~\eqref{eq_differencemethod}] based on $\bm{\kappa}$. (g)-(l) Predictions $\hat{U}$ and corresponding divergence $\partial \delta U/\partial U$ of (g),(j) a DNN and (h),(k) a linear model, as well as (i),(l) the indicator $\Delta \bar{x}$ [Eq.~\eqref{eq_differencemethod}] based on $\abs{\bm{\mathcal{F}}_{0}}$. The degree of red in (a)--(c) and (g)--(i) denotes an increasingly positive value of the respective indicator for phase transitions; (d)--(f) and (j)--(l) ground-state configurations $\bm{w}_{0}$ visualized using the same color scale as for the points in (a)--(c), respectively. These results show that the construction of local order parameters using linear models can equally be carried out in the noise-free case and using the set of correlation functions $\bm{\kappa}$ as input.}
		\label{fig_fig3_different_inputs}
	\end{center}
\end{figure*}

This section contains complementary figures which depict the vector field arising in the prediction-based method (Fig.~\ref{fig_results_vector_field}), the inferred two-dimensional phase diagram using the mean-based method with $\abs{\bm{\mathcal{F}}_{0}}$ (as opposed to correlation functions) as input (Fig.~\ref{fig_results_inputdifference_noisefree_noisy}), and the analysis of the line-scan at $\rho=63/400\approx 0.16$ for different types of inputs in the noise-free and noisy case (Fig.~\ref{fig_fig3_different_inputs}). 

\end{document}